\documentclass[12pt,letterpaper]{article}
\usepackage{amsmath,amssymb,array,calc,rotating,epsfig,psfrag, amscd, datetime}
\usepackage{color}
\usepackage[
      colorlinks=true,
      urlcolor=blue,
      filecolor=blue,
      citecolor=red,
      pdfstartview=FitV,
       bookmarksopen=true
      ]{hyperref}
\usepackage[left=2cm,top=1cm,right=3cm,nohead]{geometry}

\newcommand\fft[2]{\frac{#1}{#2}}
\newcommand\ft[2]{{\textstyle\frac{#1}{#2}}}
\newcommand{\btop}[2]{\genfrac{[}{]}{0pt}{}{\,#1\,}{\,#2\,}}

\newcommand{\nc}{\newcommand}

\nc{\ra}{\rightarrow}
\nc{\lra}{\leftrightarrow}
\nc{\Ra}{\Rightarrow}
\nc{\LRa}{\Leftightarrow}
\nc{\blp}{{\big (}}
\nc{\brp}{{\big )}}
\nc{\Blp}{{\Big (}}
\nc{\Brp}{{\Big )}}
\nc{\bglp}{{\bigg (}}
\nc{\bgrp}{{\bigg )}}
\nc{\Bglp}{{\Bigg (}}
\nc{\Bgrp}{{\Bigg )}}
\nc{\slb}{{\rm [}}
\nc{\srb}{{\rm ]}}
\nc{\bslb}{{\rm \big [}}
\nc{\bsrb}{{\rm \big ]}}
\nc{\Bslb}{{\rm \Big [}}
\nc{\Bsrb}{{\rm \Big ]}}
\def\al{\alpha}

\def\eps{\epsilon}
\nc{\veps}{\varepsilon}
\def\gam{\gamma}

\def\lam{\lambda}
\def\om{\omega}
\nc{\vphi}{\varphi}
\def\tha{\theta}
\def\sig{\sigma}

\def\Gam{\Gamma}
\def\Lam{\Lambda}
\def\Om{\Omega}
\def\Sig{\Sigma}

\nc{\myvspace}{\rule[-1em]{0pt}{2.5em}}
\nc{\bea}{\begin{eqnarray}}
\nc{\eea}{\end{eqnarray}}
\nc{\be}{\begin{equation}}
\nc{\ee}{\end{equation}}
\nc{\barr}{\begin{array}}
\nc{\earr}{\end{array}}
\nc{\cA}{{\cal A}}
\nc{\cB}{ \cal B}

\def\cD{{\cal D}}

\nc{\cF}{{\cal F}}
\nc{\cG}{{\cal G}}

\def\cI{{\cal I}}

\nc{\cL}{{\cal L}}
\nc{\cM}{{\cal M}}
\def\N{{\cal N}}

\nc{\cQ}{{\cal Q}}
\nc{\cR}{{\cal R}}

\def\cV{{\cal V}}
\def\cV{{\cal V}}

\def\cZ{{\cal Z}}
\nc{\cQd}{\cQ^{\dagger}}
\nc{\cRd}{\cR^{\dagger}}
\nc{\BB}{{\mathbb B}}
\nc{\CC}{{\mathbb C}}
\nc{\DD}{{\mathbb D}}
\nc{\EE}{{\mathbb E}}
\nc{\FF}{{\mathbb F}}
\nc{\GG}{{\mathbb G}}
\nc{\HH}{{\mathbb H}}
\nc{\JJ}{{\mathbb J}}
\nc{\RR}{{\mathbb R}}
\nc{\PP}{{\mathbb P}}
\nc{\QQ}{{\mathbb Q}}
\nc{\ZZ}{{\mathbb Z}}
\nc{\calone}{{\mathbb 1}}

\nc{\half}{\frac{1}{2}}
\nc{\qrt}{\frac{1}{4}}
\nc{\del}{\partial}

\nc{\delbar}{\bar\partial}
\nc{\thalf}{\frac{t}{2}}
\nc{\Spin}{\operatorname{Spin}}
\nc{\SO}{\operatorname{SO}}

\nc{\Sp}{{\rm Sp}}
\nc{\com}[2]{{ \left[ #1, #2 \right] }}
\nc{\acom}[2]{{ \left\{ #1, #2 \right\} }}
\nc{\rr}{\rightarrow}
\nc{\p}{\partial}
\nc{\LT}{{\LL_\T}}
\nc{\Tr}{{\rm Tr}}
\nc{\tr}{{\rm tr}}
\def\com#1#2{{ \left[ #1, #2 \right] }}
\def\acom#1#2{{ \left\{ #1, #2 \right\} }}
\nc{\Adag}{A^{\dagger}}
\nc{\AdagI}{A^{\dagger I}}
\nc{\AdagJ}{A^{\dagger J}}
\nc{\AdagK}{A^{\dagger K}}
\nc{\AdagL}{A^{\dagger L}}
\nc{\AdagM}{A^{\dagger M}}
\nc{\Bdag}{B^{\dagger}}
\nc{\BdagI}{B^{\dagger}_I}
\nc{\BdagJ}{B^{\dagger}_J}
\nc{\BdagK}{B^{\dagger}_K}
\nc{\BdagL}{B^{\dagger}_L}
\nc{\BdagM}{B^{\dagger}_M}
\nc{\Cdag}{C^{\dagger}}
\nc{\CdagI}{C^{\dagger I}}
\nc{\CdagJ}{C^{\dagger J}}
\nc{\CdagK}{C^{\dagger K}}
\nc{\Ddag}{D^{\dagger}}
\nc{\DdagI}{D^{\dagger I}}
\nc{\DdagJ}{D^{\dagger J}}
\nc{\DdagK}{D^{\dagger K}}
\nc{\ttha}{\tilde{\theta}}
\nc{\ttau}{\tilde{\tau}}
\nc{\tTha}{\tilde{\Theta}}
\nc{\tphi}{\tilde{\phi}}
\nc{\tsig}{\tilde{\sig}}
\nc{\tom}{\widetilde{\om}}
\nc{\tOm}{\widetilde{\Om}}
\nc{\tlam}{\tilde{\lam}}
\nc{\tLam}{\tilde{\Lam}}
\nc{\tSig}{\widetilde{\Sig}}
\nc{\tPhi}{\tilde{\Phi}}
\nc{\tPhibar}{\ol{\tPhi}}
\nc{\tPi}{\tilde{\Pi}}
\nc{\tpsi}{\tilde{\psi}}
\nc{\tPsi}{\tilde{\Psi}}
\nc{\tgam}{\widetilde{\gam}}
\nc{\tGam}{\tilde{\Gam}}
\nc{\tzeta}{\tilde{\zeta}}
\nc{\tZeta}{\tilde{\Zeta}}
\nc{\teta}{\tilde{\eta}}
\nc{\teps}{\tilde{\eps}}
\nc{\tveps}{\tilde{\veps}}
\nc{\tEta}{\tilde{\Eta}}
\nc{\tchi}{\tilde{\chi}}
\nc{\tChi}{\tilde{\Chi}}
\nc{\txi}{\tilde{\xi}}
\nc{\tXi}{\widetilde{\Xi}}
\nc{\tnu}{\tilde{\nu}}
\nc{\tmu}{\tilde{\mu}}

\nc{\tb}{\tilde b}
\nc{\tc}{\tilde c}
\nc{\te}{\tilde e}
\nc{\tf}{\tilde f}
\nc{\tg}{\tilde g}
\nc{\ti}{\tilde i}
\nc{\tj}{\tilde j}
\nc{\tk}{\tilde k}
\nc{\tl}{\tilde l}
\nc{\tm}{\tilde m}
\nc{\tn}{\tilde n}
\nc{\tp}{\widetilde{p}}
\nc{\tq}{\widetilde{q}}
\nc{\ts}{{\tilde s}}
\nc{\tu}{{\tilde u}}
\nc{\tv}{{\tilde v}}
\nc{\tw}{{\tilde w}}
\nc{\tx}{{\tilde x}}
\nc{\ty}{{\tilde y}}
\nc{\tz}{\tilde z}
\nc{\tA}{{\widetilde A}}
\nc{\tAbar}{{\ol \tA}}
\nc{\tB}{{\widetilde B}}
\nc{\tC}{{\widetilde C}}
\nc{\tD}{{\widetilde D}}
\nc{\tE}{{\widetilde E}}
\nc{\tG}{{\widetilde G}}
\nc{\tH}{{\widetilde H}}
\nc{\tJ}{{\widetilde J}}
\nc{\tJbar}{{\ol {\tilde J}}}
\nc{\tK}{{\widetilde K}}
\nc{\tL}{{\widetilde L}}
\nc{\tM}{{\widetilde M}}
\nc{\tN}{{\widetilde N}}
\nc{\tP}{{\widetilde P}}
\nc{\tQ}{{\widetilde Q}}
\nc{\tR}{{\widetilde R}}
\nc{\tS}{\widetilde{S}}
\nc{\tF}{\tilde{{\cal F}}}
\nc{\tX}{\widetilde{X}}
\nc{\tY}{\widetilde{Y}}
\nc{\tcZ}{\tilde{\cZ}}
\nc{\tcZbar}{\ol{\tcZ}}

\nc{\hb}{\hat b}
\nc{\hc}{\hat c}
\nc{\hd}{\hat d}
\nc{\he}{\hat e}
\nc{\hf}{\hat f}
\nc{\hg}{\hat g}
\nc{\hh}{\hat h}
\nc{\hp}{\hat p}
\nc{\hr}{\hat r}
\nc{\hs}{\hat s}
\nc{\hv}{\hat v}
\nc{\hw}{\hat w}
\nc{\hx}{\hat x}
\nc{\hy}{\hat y}
\nc{\hz}{\hat z}
\nc{\zhat}{\hat z}
\nc{\hA}{\widehat{A}}
\nc{\hB}{\widehat{B}}
\nc{\hC}{\widehat{C}}
\nc{\hD}{\widehat{D}}
\nc{\hE}{\widehat{E}}
\nc{\hF}{\widehat{F}}
\nc{\hG}{\widehat{G}}
\nc{\hH}{\widehat{H}}
\nc{\hJ}{\widehat{J}}
\nc{\hK}{\widehat{K}}
\nc{\hL}{\widehat{L}}
\nc{\hM}{\widehat M}
\nc{\hN}{\widehat{N}}
\nc{\hO}{\widehat{O}}
\nc{\hP}{\widehat{P}}
\nc{\hQ}{\widehat{Q}}
\nc{\hR}{\widehat{R}}
\nc{\hS}{\widehat{S}}
\nc{\hT}{\widehat{T}}
\nc{\hU}{\widehat{U}}
\nc{\hV}{\widehat V}
\nc{\hcV}{\widehat \cV}
\nc{\hX}{\widehat X}

\nc{\ha}{\widehat \alpha}
\nc{\hphi}{\hat{\phi}}
\nc{\hkap}{\hat{\kappa}}
\nc{\hchi}{\widehat{\chi}}
\nc{\hpsi}{\hat{\psi}}
\nc{\hgam}{\hat{\gam}}
\nc{\hPhi}{\hat{\Phi}}
\nc{\hPsi}{\hat{\Psi}}
\nc{\hGam}{\hat{\Gam}}
\nc{\omhat}{\hat{\om}}
\nc{\hOm}{\widehat{\Om}}

\nc{\w}{\wedge}

\nc{\vb}{\vec b}
\nc{\vc}{\vec c}
\nc{\vd}{\vec d}
\nc{\ve}{\vec e}
\nc{\vf}{\vec f}
\nc{\vg}{\vec g}
\nc{\vh}{\vec h}
\nc{\vp}{\vec p}
\nc{\vq}{\vec q}
\nc{\vr}{\vec r}
\nc{\vs}{\vec s}
\nc{\vv}{\vec v}
\nc{\vw}{\vec w}
\nc{\vx}{\vec x}
\nc{\vy}{\vec y}
\nc{\vz}{\vec z}

\nc{\vB}{\vec B}
\nc{\vC}{\vec C}
\nc{\vD}{\vec D}
\nc{\vE}{\vec E}
\nc{\vF}{\vec F}
\nc{\vG}{\vec G}
\nc{\vH}{\vec H}
\nc{\vP}{\vec P}
\nc{\vQ}{\vec Q}
\nc{\vR}{\vec R}
\nc{\vS}{\vec S}
\nc{\vV}{\vec V}
\nc{\vW}{\vec W}
\nc{\vX}{\vec X}
\nc{\vY}{\vec Y}
\nc{\vZ}{\vec Z}

\nc{\ol}{\overline}
\nc{\abar}{\ol{a}}
\nc{\bbar}{\ol{b}}
\nc{\cbar}{\ol{c}}
\nc{\dbar}{\ol{d}}
\nc{\ebar}{\ol{e}}
\nc{\fbar}{\ol{f}}
\nc{\ibar}{\ol{\imath}}
\nc{\jbar}{\ol{\jmath}}
\nc{\kbar}{\ol{k}}
\nc{\lbar}{\ol{l}}
\nc{\mbar}{\ol{m}}
\nc{\nbar}{\ol{n}}
\nc{\pbar}{\ol{p}}
\nc{\qbar}{\ol{q}}
\nc{\ubar}{\ol{u}}
\nc{\vbar}{\ol{v}}
\nc{\wbar}{\ol{w}}
\nc{\xbar}{\ol{x}}
\nc{\ybar}{\ol{y}}
\nc{\zbar}{\ol{z}}

\nc{\Abar}{\ol{A}}
\nc{\Bbar}{\ol{B}}
\nc{\Cbar}{\ol{C}}
\nc{\Dbar}{\ol{D}}
\nc{\Ebar}{\ol{E}}
\nc{\Fbar}{\ol{F}}
\nc{\Jbar}{\ol{J}}
\nc{\Kbar}{\ol{K}}
\nc{\Lbar}{\ol{L}}
\nc{\Mbar}{\ol{M}}
\nc{\Nbar}{\ol{N}}
\nc{\Pbar}{\ol{P}}
\nc{\Qbar}{\ol{Q}}
\nc{\Rbar}{\ol{R}}
\nc{\Sbar}{\ol{S}}
\nc{\Tbar}{\ol{T}}
\nc{\Ubar}{\ol{U}}
\nc{\Vbar}{\ol{V}}
\nc{\Wbar}{\ol{W}}
\nc{\Xbar}{{\overline X}}
\nc{\Ybar}{{\overline Y}}
\nc{\Zbar}{{\overline Z}}
\nc{\cZbar}{{\overline \cZ}}

\nc{\epsbar}{\ol{\epsilon}}
\nc{\lambar}{\ol{\lambda}}
\nc{\kapbar}{\ol{\kappa}}
\nc{\zetabar}{\ol{\zeta}}
\nc{\Zetabar}{\ol{\Zeta}}
\nc{\taubar}{\ol{\tau}}
\nc{\Taubar}{\ol{\Tau}}
\nc{\psibar}{\ol{\psi}}
\nc{\Psibar}{\ol{\Psi}}
\nc{\phibar}{\ol{\phi}}
\nc{\Phibar}{\ol{\Phi}}
\nc{\chibar}{\ol{\chi}}
\nc{\mubar}{\ol{\mu}}
\nc{\nubar}{\ol{\nu}}
\nc{\rhobar}{\ol{\rho}}
\nc{\ombar}{\ol{\om}}
\nc{\Ombar}{\ol{\Om}}
\nc{\Deltabar}{\ol{\Delta}}
\nc{\Thetabar}{\ol{\Theta}}
\nc{\xibar}{\ol{\xi}}
\nc{\Xibar}{\ol{\Xi}}

\nc{\Dthbar}{\ol{\rm D3}}

\nc{\gdot}{\dot{g}}
\nc{\xdot}{\dot{x}}
\nc{\ydot}{\dot{y}}
\nc{\phidot}{\dot{\phi}}
\nc{\sinp}{s_{\phi}}
\nc{\cosp}{c_{\phi}}
\nc{\tanp}{t_{\phi}}
\nc{\spone}{s_{\phi_1}}
\nc{\cpone}{c_{\phi_1}}
\nc{\tpone}{t_{\phi_1}}
\nc{\sptwo}{s_{\phi_2}}
\nc{\cptwo}{c_{\phi_2}}
\nc{\tptwo}{t_{\phi_2}}
\nc{\spth}{s_{\phi_3}}
\nc{\cpth}{c_{\phi_3}}
\nc{\tpth}{t_{\phi_3}}
\nc{\calp}{c_{\al}}
\nc{\salp}{s_{\al}}

\nc{\csch}{{\rm csch}}
\nc{\sech}{{\rm sech}}

\nc{\cothzlami}{\coth(z-\lam_i)}
\nc{\coshzlami}{\cosh(z-\lam_i)}
\nc{\sinhzlami}{\sinh(z-\lam_i)}

\nc{\cothzlamj}{\coth(z-\lam_j)}
\nc{\coshzlamj}{\cosh(z-\lam_j)}
\nc{\sinhzlamj}{\sinh(z-\lam_j)}

\nc{\cothlamij}{\coth(\lam_i-\lam_j)}
\nc{\coshlamij}{\cosh(\lam_i-\lam_j)}
\nc{\sinhlamij}{\sinh(\lam_i-\lam_j)}
\nc{\bah}{{\mathbf {\hat{A}}}}
\nc{\bX}{{\mathbf X}}
\nc{\ba}{{\bf a}}
\nc{\bb}{{\bf b}}
\nc{\bc}{{\bf c}}
\nc{\bd}{{\bf d}}
\nc{\bg}{{\bf g}}
\nc{\bk}{{\bf k}}
\nc{\bl}{{\bf l}}
\nc{\bm}{{\bf m}}
\nc{\bn}{{\bf n}}
\nc{\bo}{{\bf o}}
\nc{\bp}{{\bf p}}
\nc{\bq}{{\bf q}}
\nc{\br}{{\bf r}}
\nc{\bs}{{\bf s}}
\nc{\bt}{{\bf t}}
\nc{\bu}{{\bf u}}
\nc{\bv}{{\bf v}}
\nc{\bw}{{\bf w}}
\nc{\bx}{{\bf x}}
\nc{\by}{{\bf y}}
\nc{\bz}{{\bf z}}
\nc{\bom}{{\bf \om}}
\nc{\bombar}{{\mathbf \ombar}}
\nc{\bPhi}{{\bf \Phi}}

\nc{\rma}{{\rm a}}
\nc{\rmb}{{\rm b}}
\nc{\rmc}{{\rm c}}
\nc{\rmd}{{\rm d}}
\nc{\rmg}{{\rm g}}
\nc{\rk}{{\rm k}}
\nc{\rml}{{\rm l}}
\nc{\rmm}{{\rm m}}
\nc{\rmn}{{\rm n}}
\nc{\rmo}{{\rm o}}
\nc{\rmp}{{\rm p}}
\nc{\rmq}{{\rm q}}
\nc{\rmr}{{\rm r}}
\nc{\rms}{{\rm s}}
\nc{\rmt}{{\rm t}}
\nc{\rmu}{{\rm u}}
\nc{\rmv}{{\rm v}}
\nc{\rmw}{{\rm w}}
\nc{\rmx}{{\rm x}}
\nc{\rmy}{{\rm y}}
\nc{\rmz}{{\rm z}}
\nc{\Ffour}{{F^{(4)}}}
\nc{\Ffive}{{F^{(5)}}}

\nc{\dal}{\dot{\al}}
\nc{\thadot}{\dot{\tha}}
\nc{\thab}{\bar{\theta}}
\nc{\thal}{\theta^{\al}}
\nc{\thdal}{\bar{\theta}^{\dal}}

\nc{\thsigthm}{\tha \sigma^m \thab}
\nc{\thsigthn}{\tha \sigma^n \thab}

\nc{\Dal}{D_{\al}}
\nc{\Ddal}{\bar{D}_{\dal}}
\nc{\CDal}{{\cal D}_{\al}}
\nc{\CDdal}{\bar{\cal D}_{\dal}}

\nc{\eq}[1]{(\ref{#1})}
\nc{\non}{\nonumber}

\nc{\equ}{{\rm eq}}
\def\Im{{\rm Im ~}}
\def\IOm{{\rm \Om_I}}

\def\Re{{\rm Re ~}}
\def\ROm{{\rm \Om_R}}

\nc{\vol}{{\rm vol}}
\nc{\Ainf}{A_{\infty}}
\nc{\End}{{\rm End}}
\nc{\Ext}{{\rm Ext}}
\nc{\IIB}{{\rm IIB}}
\nc{\Ad}{{\rm Ad}}
\nc{\IIA}{{\rm IIA}}
\nc{\AdS}{{\rm AdS}}
\nc{\CFT}{{\rm CFT}}
\nc{\Dslash}{\ensuremath \raisebox{0.025cm}{\slash}\hspace{-0.32cm} D}
\nc{\cDslash}{\ensuremath \raisebox{0.025cm}{\slash}\hspace{-0.32cm} \cD}
\nc{\low}{\leftarrow}
\nc{\row}{\rightarrow}
\nc{\no}{\!:\!\!}
\nc{\ointdz}{\oint\frac{dz}{2\pi i}}
\nc{\ointdzone}{\oint\frac{dz_1}{2\pi i}}
\nc{\ointdztwo}{\oint\frac{dz_2}{2\pi i}}
\nc{\ointdzb}{\oint\frac{d\zbar}{2\pi i}}
\nc{\ointdzbone}{\oint\frac{d\zbar_1}{2\pi i}}
\nc{\ointdzbtwo}{\oint\frac{d\zbar_2}{2\pi i}}
\nc{\dz}{\frac{dz}{2\pi i}}
\nc{\dzb}{\frac{d\zbar}{2\pi i}}
\nc{\bpm}{\begin{pmatrix}}
\nc{\epm}{\end{pmatrix}}
 \nc{\bitem}{\begin{itemize}}
 \nc{\eitem}{\end{itemize}}

\definecolor{cardinal}{rgb}{0.6,0,0}
\definecolor{darkgreen}{rgb}{0,0.5,0}
\definecolor{golden}{rgb}{0.92, 0.7, 0}
\definecolor{midnight}{rgb}{0, 0, 0.5}
\definecolor{darkblue}{rgb}{0.2, 0, 0.8}

\begin{document}

\begin{center}
\hfill     MCTP-11-41
\vskip 2 cm
{\Large \bf  On $\mathcal N = 2$ Truncations of IIB on $T^{1,1}$ }
\vskip 1 cm

Nick Halmagyi$^{*}$, James T.~Liu$^{\dagger}$ and Phillip Szepietowski$^{\ddagger}$ \\
 \vskip0.5cm

$^{*}$\textit{Laboratoire de Physique Th\'eorique et Hautes Energies,\\
Universit\'e Pierre et Marie Curie, CNRS UMR 7589, \\
F--75252 Paris Cedex 05, France}\\
\vskip0.5cm
$^{*}$\textit{Center for the Fundamental Laws of Nature \\
Harvard University, Cambridge, MA 02138, USA} \\
\vskip 0.5cm
$^{\dagger}$\textit{Michigan Center for Theoretical Physics, Randall Laboratory of Physics,\\
The University of Michigan, Ann Arbor, MI 48109--1040, USA}\\
\vskip0.5cm
$^{\ddagger}$\textit{Department of Physics, University of Virginia,\\
Box 400714, Charlottesville, VA 22904, USA} \\
\let\thefootnote\relax\footnotetext{

  $^{*}$halmagyi@lpthe.jussieu.fr

  $^{\dagger}$jimliu@umich.edu

  $^{\ddagger}$pgs8b@virginia.edu}
\end{center}

\begin{abstract}
We study the $\N=4$ gauged supergravity theory which arises from the consistent truncation of IIB supergravity on the coset $T^{1,1}$. We analyze three $\N=2$ subsectors and in particular we clarify the relationship between true superpotentials for gauged supergravity and certain fake superpotentials which have been widely used in the literature. We derive a superpotential for the general reduction of type I supergravity on $T^{1,1}$ and this together with a certain solution generating symmetry is tantamount to a superpotential for the baryonic branch of the Klebanov-Strassler solution.
\end{abstract}

\section{Introduction}

Starting with the work \cite{Klebanov:1998hh}, the study of type IIB supergravity on the conifold has given rise to much progress in gauge/gravity duality. In particular, it provides an example of a gravity dual to a non-conformal, four dimensional field theory with minimal supersymmetry \cite{Klebanov:2000hb}. This background, known as the warped deformed conifold, can be used to model the local geometry of a flux compactification \cite{Giddings:2001yu}. In the current work, following \cite{Bena:2010pr, Cassani:2010na}, we study the gauged supergravity theory which arises from Kaluza-Klein reduction of IIB supergravity on the coset $T^{1,1}$.

The Kaluza-Klein reduction of ten and eleven dimensional supergravity to lower dimensional gauged supergravity theories has a rich history. In particular there has been much attention applied to the case of reduction on spheres down to maximally supersymmetric gauged supergravity \cite{deWit:1984nz, Nastase:1999cb}. Another route to deriving lower dimensional gauged supergravity theories is to use a set of globally defined fundamental forms on the internal manifold which close  under exterior derivative and wedge product. This technique has been used for nearly K\"ahler manifolds \cite{KashaniPoor:2007tr}, cosets \cite{House:2005yc, Cassani:2009ck, Bena:2010pr, Cassani:2010na}, Sasaki-Einstein manifolds \cite{Gauntlett:2009zw, Gauntlett:2010vu, Liu:2010sa, Skenderis:2010vz, Cassani:2010uw} and also more general flux backgrounds in \cite{Donos:2010ax, Colgain:2011ng, Danckaert:2011ju}. Additionally, recent progress has been made exploring the fermion sector of these reductions \cite{Bah:2010yt,Bah:2010cu,Liu:2010pq,Liu:2011dw}.

The current work synthesizes aspects of the Kaluza-Klein reduction of IIB supergravity on $T^{1,1}$ performed in \cite{Bena:2010pr, Cassani:2010na} that retains just the singlet sector under the global symmetries of $T^{1,1}$. In fact similar reductions (restricted to just the scalar sector) were employed to derive the warped deformed conifold solution \cite{Klebanov:2000hb,Klebanov:2000nc} (and used in many other scenarios as well \cite{Gubser:2001eg, Papadopoulos:2000gj, PandoZayas:2000sq, PandoZayas:2001iw, Borokhov:2002fm, Bena:2009xk}), where  a one-dimensional action was derived and a superpotential found from which one can compute the scalar potential. This superpotential was then used to facilitate the supersymmetry analysis and thus bypass using ten dimensional spinors directly. In more recent work \cite{Bena:2010pr, Cassani:2010na}, it was found that there exists a supersymmetric Kaluza-Klein reduction on $T^{1,1}$ down to five dimensional $\N=4$ gauged supergravity (generalizing the work on Sasaki-Einstein manifolds \cite{Gauntlett:2010vu, Liu:2010sa, Skenderis:2010vz,Cassani:2010uw}) from which all these one dimensional models can be obtained by additional reduction on $\RR^{1,3}$ and some further truncation of the fields.

The advantages of performing a rigorous supersymmetric reduction, thus including higher form fields and not just the scalar sector, are manyfold. It allows for a simple yet rigorous analysis of supersymmetric solutions, it allows one to consider solutions with non-trivial profiles for form fields relevant for AdS/CMT \cite{Gauntlett:2009dn, Gauntlett:2009bh}, and it also helps to characterize which gauged supergravity theories can be obtained from string theory.

One goal of the current work is to develop the $\N=2$ five dimensional gauged supergravity theories which are relevant for studying the physics of the warped deformed conifold solution and its relatives. One such $\N=2$ theory is obtained by truncating to modes which are even under a particular $\ZZ_2$ symmetry $\cI$ which will be explained in section \ref{sec:bh}. Within this $\cI$ invariant truncation there exists a superpotential $W_{KS}$ which has been known for some time \cite{Papadopoulos:2000gj}. But as we will see, $W_{KS}$ is in fact a {\it fake} superpotential even though the theory is supersymmetric. It was essentially noticed in \cite{Kuperstein:2003yt} that from $W_{KS}$ one can derive a solution for fluxes on the warped deformed conifold which are known from ten dimensional analysis \cite{Grana:2001xn} to be non-supersymmetric. The analysis we perform resolves this seeming discrepancy since we can identify precisely how $W_{KS}$ fails to be a true superpotential of the theory. We can then characterize which fluxes are in fact supersymmetric on the warped deformed conifold.

While there have been superpotentials provided for the solution of \cite{Chamseddine:1997nm, Maldacena:2000yy} and also \cite{Klebanov:2000hb}, it has been an open problem for some time to provide a superpotential for the interpolating solution of \cite{Butti:2004pk}. In section \ref{sec:NS} we study the sector of the $\N=4$ theory corresponding to retaining just $(g_{MN}, \phi, F_{3})$ which we will call the NS-sector truncation%
\footnote{We are abusing notation here since we keep $F_{3}$ and not $H_{3}$. But these are related by S-duality}.
Importantly, this sector retains $\cI$-even and $\cI$-odd modes, and we derive a superpotential for this truncation. Using the TST duality transformation of \cite{Maldacena:2009mw}, from any solution of the NS-truncation one can generate a family of solutions which lie within the $\N=4$ theory. As such, our superpotential can be considered a superpotential for the baryonic branch of the warped deformed conifold.

The organization of the rest of this paper is as follows. In the following section we lay the ground work for the $\mathcal N = 4$ reduction of IIB on $T^{1,1}$. In addition we analyze the duality group of the $\mathcal N = 4$ theory, notably finding the embedding of the $SL(2,\RR)$ of IIB supergravity within the $\mathcal N=4$ scalar coset. In Section \ref{sec:N2gs} we review some relevant material on five-dimensional $\mathcal N = 2$ gauged supergravity coupled to vector and hyper multiplets. We also include a discussion on the existence of superpotentials, real and fake, and their relation to solutions of the BPS domain wall equations. In Section \ref{sec:N2truncs} we provide the relevant details of three truncations of the $\mathcal N = 4$ theory to $\mathcal N = 2$ gauged supergravity. We analyze the conditions imposed by supersymmetry and present superpotentials for each truncation. Finally, in Section \ref{sec:conc} we conclude with some remarks on the pitfalls and advantages of superpotential techniques. By studying a specific solution on the warped deformed conifold we detail precisely the way in which solutions found from a fake superpotential can end up being, in fact, non-supersymmetric. Additionally, we remark on potential future work towards understanding relations between the current work and solution generating techniques such as the TST transformation in string theory.

For sake of clarity we have relegated many important details of the $\mathcal N = 2$ truncations to Appendices~\ref{app:BettiHyper}, \ref{app:BettiVector} and \ref{app:NS}. Specifically, for each truncation we include a detailed description of the scalar coset manifolds and the coordinate transformations which lift the coset coordinates to IIB supergravity fields. In addition we present the reduction of the IIB fermion variations, which we find to be consistent with the scalar coset structure, as expected.  Finally, Appendix~\ref{app:map} summarizes some differences in convention between
the present work and Refs.~\cite{Bena:2010pr,Liu:2011dw} concerning the $T^{1,1}$ reduction.

\section{$\mathcal N=4$ gauged supergravity from IIB on $T^{1,1}$} \label{sec:N4}

The consistent truncation of IIB supergravity on $T^{1,1}$ was performed in
\cite{Bena:2010pr, Cassani:2010na}, and the resulting theory is described by gauged
$\mathcal N=4$ supergravity in five dimensions coupled to three vector multiplets. Since
this is the starting point for the further $\mathcal N=2$ truncations, we first review this
construction, establish notation and derive the action of the IIB $SL(2,\RR)$ symmetry
on the gauged supergravity theory.

The bosonic field content of IIB supergravity consists of the metric, IIB axi-dilaton
$\tau=a+ie^{-\phi}$, three-forms $F_3^i$ ($i=1,2$) and RR five-form $\tilde F_5$.
The ten dimensional metric is reduced according to
\bea
ds_{10}^2&=&  e^{2u_3-2u_1} ds_5^2 + e^{2u_1+2u_2} E'_1 \Ebar'_1 +e^{2u_1-2u_2} E'_2 \Ebar'_2+ e^{-6u_3-2u_1} E_5 E_5 \label{metansatz},
\eea
where
\bea
E_1=\frac{1}{\sqrt{6}}\blp  \sig_1+i\sig_2 \brp, &&E_2=\frac{1}{\sqrt{6}}\blp \Sig_1+i\Sig_2 \brp ,
\nonumber\\
E'_1=E_1\,, && E'_2= E_2+v \Ebar_1,\nonumber\\
E_5 = g_5+A_1, && g_5=\frac{1}{3}\blp \sig_3+\Sig_3\brp,
\eea
and the $SU(2)$-invariant one forms satisfy $d\sig_i=\half\eps_{ijk} \sig_j\w \sig_k$ and $d\Sig_i=\half\eps_{ijk} \Sig_j\w \Sig_k$.  This follows from writing $T^{1,1}$ as $U(1)$ bundled over $\mathbb P^1\times\mathbb P^1$.  In particular, the $U(1)$ structure may be described by the invariant forms
\begin{eqnarray}
J_1=\fft{i}2E_1\wedge\bar E_1,\qquad J_2=\fft{i}2E_2\wedge\bar E_2,\qquad
\Omega=E_1\wedge E_2.
\end{eqnarray}
The reduction of the metric yields three real five-dimensional scalars $(u_1,u_2,u_3)$,
one complex scalar $v$, and a $U(1)$ gauge field $A_1$ with field strength $F_2=dA_1$.

We adopt a mixed notation with respect to \cite{Liu:2011dw} and \cite{Bena:2010pr} for the IIB forms which makes the $SL(2,\mathbb R)$ invariance explicit.  The differences in notation are summarized
in Appendix~\ref{app:map}.  For the three-forms, we expand the two form potentials as
\begin{equation}
B_2^i=b_2^i+b_1^i\wedge E_5+c_0^iJ_++e_0^iJ_-+2\Re(b_0^i\Omega),
\end{equation}
where $J_\pm=J_1\pm J_2$, and write
\begin{equation}
F_3^i=dB_2^i+j_0^iJ_-\wedge E_5,
\end{equation}
where $j_0^i$ are the charges coming from topological flux on the $S^3\subset T^{1,1}$.  Explicitly, for the three forms, we have
\begin{equation}
F_3^i=g_3^i+g_2^i\wedge E_5+(g_1^i+h_1^i)\wedge J_1
+(g_1^i-h_1^i)\wedge J_2+j_0^iJ_-\wedge E_5+2\Re[f_1^i\wedge\Omega
+f_0^i\Omega\wedge E_5],
\end{equation}
where
\begin{eqnarray}\label{eq:threeformfields}
&&g_3^i=db_2^i-b_1^i\wedge F_2,\kern4.2em g_2^i=db_1^i,\kern4em
g_1^i=dc_0^i-2b_1^i \equiv Dc_0^i ,\nonumber\\
&&h_1^i=de_0^i-j_0^iA_1 \equiv De_0^i,\qquad f_1^i=db_0^i-3ib_0^i A_1 \equiv Db_0^i,\qquad f_0^i=3ib_0^i.
\end{eqnarray}
The three-forms contribute two $SL(2,\mathbb R)$ doublets of real scalars $(c_0^i,e_0^i)$,
one doublet complex scalar $b_0^i$, one doublet of $U(1)$ gauge fields $b_1^i$ with field strength
$g_2^i=db_1^i$ and one doublet two-form potential $b_2^i$. Alternatively, one may define
the complex three-form field strength
\be
\frac{1}{\sqrt{\tau_2}}G_3 = v_i F^i_3 = \frac{1}{\sqrt{\tau_2}}\left(F^2_3 - \tau F_3^1\right),
\ee
where we have introduced the $SL(2,\mathbb R)$ vielbein $v_i$. However, we will always use a notation that leaves the $SL(2,\mathbb R)$ structure explicit.

The five-form field strength can be expanded in the basis
\bea
\widetilde F_5&=&(1+*)[e^Z J_1\wedge J_2 \wedge E_5+ K_1 \wedge J_1\wedge J_2
+K_{21}\wedge J_1\wedge E_5\nonumber\\
&&\kern3.2em
+K_{22}\wedge J_2\wedge E_5+2\Re(L_2\wedge\Omega\wedge E_5)].
\eea
The Bianchi identity $d\tilde F_5=\fft12\epsilon_{ij}F_3^i\wedge F_3^j$ yields the
constraints
\bea
e^Z &=& Q - 6i\epsilon_{ij}(b_0^i\bar{b_0^j}-\bar{b_0^i}b_0^j) + \epsilon_{ij}(j_0^i e_0^j - j_0^j e_0^i)\,, \non \\
K_1 &=& Dk + 2\epsilon_{ij}[b_0^i \bar{Db_0^j} + \bar b_0^i Db_0^j]- \epsilon_{ij}e_0^ih_1^j\,, \non \\
K_{21} &=& Dk_{11} + \ft14\epsilon_{ij}g_1^i\wedge g_1^j  + \ft12\epsilon_{ij}g_1^i\wedge h_1^j\,, \non \\
K_{22} &=& Dk_{12} + \ft14\epsilon_{ij}g_1^i\wedge g_1^j  - \ft12\epsilon_{ij}g_1^i\wedge h_1^j\,,
\eea
where the covariant derivatives are defined as
\bea
Dk &=& dk - Q A_1 - 2k_{11} - 2k_{12} - \epsilon_{ij}j_0^i e_0^j A_1\,, \non \\
Dk_{11} &=& dk_{11} - \epsilon_{ij}j_0^i b_2^j\,, \non \\
Dk_{12} &=& dk_{12} + \epsilon_{ij}j_0^i b_2^j\,.
\eea
The charge $Q$ comes from mobile D3-branes.  The five-form contributes one real scalar $k$,
two one-forms $(k_{11},k_{12})$ and a complex two-form $L_2$.

In summary, the reduction of IIB supergravity on $T^{1,1}$ yields $\mathcal N=4$
supergravity coupled to three vector multiplets.  The scalar manifold is
\be
\cM_{sc}=\frac{SO(5,n_v)}{SO(5)\times SO(n_v)} \times SO(1,1),
\ee
with $n_v=3$.  As shown in \cite{Bena:2010pr, Cassani:2010na},
the $SO(1,1)$ is parameterized by $u_3$, while the remaining
$5\times n_v=5\times3$ scalars are
\begin{equation}
(u_1,u_2,c_0^i,e_0^i,k,
\tau,\bar\tau,v,\bar v,b_0^i,\bar b_0^i).
\end{equation}
Along with the scalars,
there are a total of nine-vectors:  a singlet vector $A_1$, along with $5+n_v=8$ additional
vectors transforming in the vector representation of $SO(5,n_v)$.  The latter eight
vectors correspond to the potentials
\begin{equation}
(b_1^i,k_{11},k_{12},b_2^i,L_2,\bar L_2),
\end{equation}
where the two-form potentials $b_2^i$ and $L_2$ are dual to vectors in five dimensions.

\subsection{Duality transformations}

In ungauged supergravity with a scalar manifold given by a coset $\hG/\hH$, the duality group is given by global $\hG$ transformations. These transformations act on the coset on the right, say, and are compensated by the left action of a local $\hH$ transformation which brings the coset element back to a canonical form. After gauging, only a subgroup of $\hG$ transformations remain symmetries of the theory. It is clear for $\N=4$ theories that the commutant of the gauge group $G$ in $SO(5,n_v)$ is a symmetry of the theory. But, in addition, there could be further symmetries. There is currently no understanding in general of how large the symmetry group is or how to compute it for a given gauged supergravity theory. To perform an analysis of the duality group, the embedding tensor formalism (see {\it e.g.}\ \cite{Schon:2006kz}) is quite useful since it facilitates the embedding of the gauge group into the scalar manifold in a covariant way.

As reviewed above, the $T^{1,1}$ reduction yields $\mathcal N=4$ supergravity coupled to
three $\mathcal N=4$ vector multiplets, with the scalar manifold
\cite{Bena:2010pr, Cassani:2010na}
\be
\cM_{sc}=\frac{SO(5,3)}{SO(5)\times SO(3)} \times SO(1,1).
\ee
The field content combined with the embedding tensor \cite{Schon:2006kz} completely specify the $\N=4$ supergravity. In \cite{Bena:2010pr, Cassani:2010na} the embedding tensor $(f_{MNP},\xi_{MN})$ was shown to be 
\bea
&& f_{123}=-f_{128}=f_{137}=f_{178}=2\,, \non \\
&& \xi_{23}=-\xi_{28}=\xi_{37}=\xi_{78}=-Q/\sqrt{2}\,,   \non  \\
&& \xi_{45}=-3\sqrt{2}\,,  \non \\
&& \xi_{36}=\xi_{68}=\sqrt{2} \, j_0^2\,, \non \\
&& \xi_{26}=\xi_{67}=\sqrt{2} \, j_0^1\,, \non \\
\eea
and permutations. From this we find that the gauge group $G$ is generated by
\bea
g_0&=& 2\sqrt{3}\, t_{45} + \sqrt{2}Q\, ( t_{37}+t_{78}+t_{23}-t_{28})+\sqrt{2} j_0^2 (t_{36}+t_{68})+\sqrt{2} j_0^1 (t_{26}+t_{67})\,,\non \\
g_1&=& t_{13}-t_{18} \,, \non \\
g_2&=& t_{12}-t_{17} \,, \non \\
g_3&=&  t_{37}+t_{78}+t_{23}-t_{28} \,,
\eea
where
\be
(t_{MN})_P^{\ Q} =\delta_{[M}^Q \eta_{N]P}
\ee
are the standard generators of $SO(5,3)$ and $\eta={\rm diag}\{-1,-1,-1,-1,-1,+1,+1,+1\}$.

We find that the commutant of $G$ inside $SO(5,3)$ is in general given by the following two elements
\bea
t_{45}:&&v\ra e^{i\beta}v,\ M_0\ra e^{-i\beta}M_0,\ N_0\ra e^{-i\beta}N_0\,, \nonumber\\
t_{37}+t_{28}+t_{23}-t_{28}:&& k\ra k+\beta\,,
\eea
In addition there are two more elements
\bea
t_{26}+t_{67}:&& e_0^1\ra e_0^1 +\beta,\ k\ra k+\beta e_0^2\,, \nonumber\\
t_{36}+t_{68}:&& e_0^2\ra e_0^2 + \beta,\ k\ra k+\beta e_0^1\,
\eea
 generating symmetries which are broken by the terms in the scalar potential
\be
V_{sc}\sim j_0^2 e_0^1-j_0^1 e_0^2.
\ee

This is clearly not the full duality group since for example we at least expect to find the action of the $SL(2,\RR)$ symmetry of IIB supergravity. It turns out that this $SL(2,\RR)$ lives inside the normalizer of $G$ in $SO(5,3)$. The normalizer is ten dimensional, but by explicit computation we find that the only elements which are symmetries of the scalar potential are the realization of the $SL(2,\RR)$ symmetry of IIB supergravity. We find these to be generated by
\bea
h&=&2( t_{27}-t_{38})\,,\nonumber \\
e&=&t_{28}-t_{78}+t_{23}+t_{37}\,,\nonumber  \\
f&=&t_{28}+t_{78}-t_{23}+t_{37}\,,
\eea
satisfying
\begin{equation}
\bslb e,f\bsrb=h,\quad\bslb h,e\bsrb=2e,\quad\bslb h,f\bsrb=-2f\,.
\end{equation}
With general charges $j_0^i$, the whole symmetry is broken, but with $j_0^2=0$ ($j_0^1=0$) the symmetry generated by $e\,(f)$ survives as a symmetry of the scalar potential. When $j_0^1=j_0^2$ the full $SL(2,\RR)$ is a symmetry of the theory.

It is interesting that non-trivial duality symmetries are found outside the commutator of the gauge group inside $SO(5,3)$. In Ref.~\cite{Corrado:2002wx} the gauged supergravity was studied which arises from compactification of IIB supergravity on the orbifold $S^5/\ZZ_n$. There it was found that the commutator of the gauge group $G=SU(2)\times U(1)$, inside $SO(5,2n)$ was $SU(1,n)$. This result is at odds with the discrete duality group found in \cite{Halmagyi:2004ju} which does not quite fit inside $SU(1,n)$. It is expected that the discrete duality group is a symmetry of the dual field theory at finite $N$ and this should be enhanced to the continuous group in the limit of large $N$. (See \cite{Intriligator:1998ig} for a derivation of this fact for $N=4$ SYM in four dimensions.) What we have found here is an example of duality symmetries which lie outside the commutator of the gauge group inside $SO(5,n_v)$ and it would be interesting to explore if the duality group found in \cite{Corrado:2002wx} can be extended by considering the normalizer of the gauge group in $SO(5,2n)$.

\section{Preliminaries on $\mathcal N=2$ gauged supergravity}{\label{sec:N2gs}}

Before examining the various truncations of the $\mathcal N=4$ theory, we first review some of the
salient features of $\mathcal N=2$ gauged supergravity.  In general, $\mathcal N=2$ supergravity
may be coupled to vector, tensor and hypermultiplets.  However, we will not consider tensor
multiplets, as they will not appear in any of the truncations.  As is well known, the bosonic
field content of this theory consists of the metric $g_{\mu\nu}$, $n_v+1$ vectors $A_\mu^I$ (with
$I=0,\ldots,n_v$), $n_v$ vector multiplet scalars $\phi^x$ living on a very special
manifold and $4n_h$ hyperscalars $q^X$ on a quaternionic manifold.

The bosonic $\mathcal N=2$ Lagrangian is
\bea
\mathcal L &=& R - \fft12 g_{xy}D_\mu\phi^x D^\mu\phi^y - \fft12 g_{XY}D_\mu q^X D^\mu q^Y - V  \non \\
&& - \fft14 G_{IJ}F^I_{\mu\nu}F^{J\,\mu\nu} + \fft1{24}c_{IJK}\epsilon^{\mu\nu\rho\lambda\sigma}F^I_{\mu\nu}F^J_{\rho\lambda}A^K_\sigma,
\eea
and the fermionic supersymmetry transformations are (for the gravitino, gauginos and
hyperinos)
\bea
\delta\psi_{\mu\,i}&=&\bigl[D_\mu +\ft{i}{24} X_I(\gamma_\mu{}^{\nu\rho} - 4\delta_\mu^\nu\gamma^\rho)F_{I\,\nu\rho}\bigr]\epsilon_i + \ft{i}6 X^I (P_I)_i{}^j\epsilon_j\,, \non \\
\delta\lambda^x_i &=&\bigl(-\ft{i}2\gamma\cdot D\phi^x - \ft14g^{xy}\partial_y X^I\gamma^{\mu\nu}F_{I\,\mu\nu}\bigr)\epsilon_i - g^{xy}\partial_y X^I(P_I)_i{}^j\epsilon_j\,, \non \\
\delta\zeta^A &=& f^{i\,A}_X\bigl(-\ft{i}2\gamma\cdot Dq^X + \ft12 X^I K^X_I\bigr)\epsilon_i\,.
\label{eq:susy}
\eea
The covariant derivatives are
\bea
D_\mu\phi^x &=& \partial_\mu\phi^x + A^I_\mu K_I^x(\phi^x)
\eea
for the vector multiplet scalars and
\bea
D_\mu q^X &=& \partial_\mu q^X + A^I_\mu K_I^X(q^X)
\eea
for the hypermultiplet scalars, where we have fixed the gauge coupling $g=1$. The Killing vectors $K_I^x(\phi^x)$ and $K_I^X(q^X)$ correspond to the gauging of the isometries of the very special
manifold and quaternionic manifold, respectively.

The vector multiplet scalars are given in terms of the $n_v + 1$ constrained scalars
$X^I = X^I(\phi^x)$ subject to the very special geometry constraint
\be
\fft16c_{IJK}X^IX^JX^K = 1.
\ee
Additionally, the scalar metric for the vector multiplet scalars is determined by
\bea
G_{IJ} &=& X_IX_J - c_{IJK}X^K\,, \non\\
X_I &=& \fft12 c_{IJK}X^JX^K\,, \non\\
g_{xy} &=& \partial_x X^I \partial_y X^JG_{IJ}\,.
\eea

The Killing prepotentials $(P_I)_{i}^{\ j}=P^r_I(i\sig^r)_i^{\ j}$ are determined by the Killing vectors and depend only on the hyperscalars. They satisfy
\be
 \iota_{K_I} \Om^r = dP^r_I + \eps^{rst} \om^s P^t_I\,, \label{KP1}
\ee
where $\omega^s$ is the $SU(2)$ connection, or in co-ordinates
\be
K^X_{I} \Om^r_{XY}= \nabla_Y P^r_I\,. \label{KP2}
\ee
Here $\Om^r$ are the triplet of covariantly constant two-forms on the quaternion manifold.
While this is a differential equation for the Killing prepotentials, one can solve for them algebraically by using the fact \cite{D'Auria:2001kv} that $P^r_I$ are eigenfunctions of the Laplacian
\be
\nabla^X\nabla_X P^r_{I}= -4n_h P^r_{I}\,.
\ee
We then see that
\be
P^r_I=-\frac{1}{4n_h}\nabla^X \blp K^Y_\Lam \Om^r_{XY}\brp \label{PKOm}
\ee
is a solution to \eq{KP2}. Note that the Killing prepotentials are unique only up to a local $SU(2)$ gauge transformation.
Finally, the scalar potential couples the hypermultiplet scalars to the vector multiplet scalars and is given by
\be
V = 2g^{xy}\partial_x X^I \partial_y X^J P^r_IP^r_J -\fft43 P^rP^r + \fft12g_{XY} K^X K^Y,
\label{eq:spot}
\ee
where $P^r = X^I P_I^r$.  For convenience, we will often denote $P^r$ as an $SU(2)$ vector,
namely $\vec P = (P^1,P^2,P^3)$.

\subsection{Real and fake $\mathcal N=2$ superpotentials}

As we will discuss in the following subsection, the construction of BPS solutions to gauged
supergravity is often based on solving first order equations constructed
from the $\mathcal N=2$ superpotential.  In the absence of hypermatter, where a rigid
$U(1)$ is gauged in $SU(2)$, the Killing prepotentials are all aligned, say in the $r=3$ direction.
In this case, the superpotential is given by $W=X^IP_I^3$, and the scalar potential is
determined in the usual manner by
\begin{equation}
V=2g^{xy}\partial_xW\partial_yW-\fft43W^2\,,
\label{eq:VfromW}
\end{equation}
in perfect agreement with (\ref{eq:spot})

It is often assumed that a superpotential will continue to exist when hypermatter is
included.  However, comparing the actual potential (\ref{eq:spot}) with the expression
(\ref{eq:VfromW}) indicates a couple of differences.  Firstly, the gauging of isometries of
the quaternion manifold gives rise to an additional contribution $\fft12g_{XY}K^XK^Y$
to the potential. Secondly, the first term of (\ref{eq:spot}) only agrees with the first term
of (\ref{eq:VfromW}) for rigid $P_I^r$, since $W$ was obtained by aligning $P_I^r$ along $r=3$.
Nevertheless, it is possible to come close by defining a superpotential \cite{Ceresole:2001wi}
\be
W= \sqrt{P^rP^r}\,.
\label{eq:sqrtW}
\ee
and using the relation
\be
V = 2g^{\Lambda\Sigma}\partial_\Lambda W\partial_\Sigma W - \fft43 W^2 \,, \label{VWW}
\ee
where $\Lambda,\Sigma$ run over both vector multiplet and hypermultiplet
scalars.  But in order for this relation to work, a further constraint on the phase of $P^r$
must hold off-shell:
\be
\del_x Q^r=0\,, \label{dQVWW}
\ee
where
\be
P^r = W Q^r\,.
\ee
This condition is essentially a requirement that any $U(1)$ component that is being gauged
inside $SU(2)$ must be rigid as a function of the vector multiplet scalars.  This condition
will hold if, {\it e.g.}, the gauging of $SU(2)$ is aligned with $r=3$.  However, this is a special
case, and we will find explicit examples below where this constraint is in fact not satisfied
off-shell.

Even when a particular gauging does not admit a superpotential, in some cases it is nevertheless
possible to find a {\it fake} superpotential that reproduces the correct scalar potential using
the relation (\ref{VWW}).  In this case, one can still write down first order equations for
domain wall solutions.  However, there is no guarantee that such solutions are actually
supersymmetric; only examination of the true Killing spinor equations obtained from
(\ref{eq:susy}) will indicate whether the BPS conditions are satisfied or not.  In practice,
most solutions obtained in this fashion are supersymmetric.  However, we are not aware of
a general principle governing the existence of a fake superpotential nor determining when
the resulting solution is supersymmetric.

An alternate approach to obtaining BPS solutions in the absence of a true superpotential is
to nevertheless use the square-root superpotential (\ref{eq:sqrtW}) to derive a set of first order
equations.  In general, the result of solving this system may not satisfy the true equations of
motion.  However, once we impose the constraint (\ref{dQVWW}), the background is then
guaranteed to be a solution to the equations of motion as well as BPS.  In fact, all BPS
domain wall solutions may be obtained in this fashion.  We explore this in a bit more detail below.

\subsection{BPS domain-wall equations}

A particularly interesting class of solutions in gauged supergravity are BPS domain walls. The domain wall ansatz is given by the five-dimensional metric
\be
ds_5^2 = dr^2 + a(r)^2\eta_{\mu\nu}dx^\mu dx^\nu\,,
\ee
and is supported by scalar fields that depend only on $r$.  The vector fields vanish because of
the isometry.
Given this ansatz, it was shown in \cite{Ceresole:2001wi} that the BPS equations are given by
\bea
\fft{1}{a}\fft{da(r)}{dr} &=& \pm \fft{1}{3}W\,,  \label{eq:warpBPS}  \\
\fft{d\phi^\Lambda}{dr} &=& \mp 2 g^{\Lambda\Sigma}\partial_\Sigma W\,, \label{eq:scBPS}  \\
\partial_x Q^r &=& 0\,. \label{eq:Qr}
\eea
The curious equation here is \eq{eq:Qr} which is not a standard BPS flow equation but is equivalent to the constraint encountered above in \eq{dQVWW}.

It is worthwhile to formally analyze the constraint (\ref{eq:Qr}) a little further.  Recalling that
$\vec Q=\vec P/|\vec P|$, we find that this constraint is equivalent to
\be
\vec{P}\times(\vec{P}\times\partial_x\vec{P}) = 0\qquad
\Rightarrow \qquad \vec{P}\times\partial_x\vec{P}= 0\,.
\ee
Moreover multiplying this expression by $G_{IK}\partial_xX^K$ and using the special geometry relation \cite{Behrndt:1998jd}
\be
G_{IK}\partial^xX^K\partial_x X^J = \delta_I{}^J - X_I X^J\,,
\ee
we see that
\be
0=G_{IK}\partial^xX^K\vec{P}\times\partial_x\vec{P}
= \vec{P}\times\vec{P}_I,
\ee
As a result, the constraint implies that
\be\label{eq:cross}
\vec{P}\times\vec{P}_I = 0.
\ee
We now conclude that the only way to satisfy (\ref{eq:Qr}) is to have $\vec{P}$ identically zero or to have every nonzero $\vec{P}_I$ lie along the same direction in $SU(2)$, with possibly an arbitrary number of the $\vec{P}_I$ vanishing. An equivalent statement is to say that all cross products between any two prepotentials must vanish
\be
\vec{P}_I\times\vec{P}_J = 0\,. \label{crossvanish}
\ee
This demonstrates that the square-root superpotential (\ref{eq:sqrtW}) can be used to obtain
BPS domain wall solutions when combined with the constraint that all prepotentials are
parallel in $SU(2)$ space.  This constraint was observed in \cite{Ceresole:2001wi}
at fixed points of the domain wall flow.  However, here we have shown that the parallel
constraint must hold along all points of the supersymmetric flow. Additionally, this constraint was discussed in \cite{Behrndt:2000km}, however it was not recognized as a necessary condition of the BPS equations.

\section{The truncations to $\mathcal N=2$ gauged supergravity}\label{sec:N2truncs}

It is generally useful to restrict our attention to $\mathcal N=2$ subsectors of the full theory
when looking for BPS solutions.  This is because we may then apply the well-studied
flow equations (\ref{eq:warpBPS}) and (\ref{eq:scBPS}) along with all its associated machinery.
Starting from $\mathcal N=4$ supergravity coupled to three vector multiplets, the truncation
to $\mathcal N=2$ proceeds by removing the massive $\mathcal N=2$ gravitino multiplet.
Since the $\mathcal N=4$ gravity multiplet  reduces to a gravity multiplet coupled to a
gravitino and a vector multiplet, and each $\mathcal N=4$ vector reduces to a vector
multiplet and a hypermultiplet, the decomposition gives four vector multiplets and three
hypermultiplets.  However, the massive gravitino multiplet will eat two vector multiplets, so
upon truncation we are limited to at most two vector multiplets and three hypermultiplets
\cite{Cassani:2010na}.

Compared to the reduction on a generic Sasaki-Einstein manifold, the reduction on $T^{1,1}$
yields one additional $\mathcal N=4$ vector multiplet, denoted the Betti vector multiplet
in \cite{Cassani:2010na}.  Furthermore, Ref.~\cite{Cassani:2010na} considered two
truncations to $\mathcal N=2$.  The first retains the $\mathcal N=2$ Betti hypermultiplet,
and gives rise to a total of one vector multiplet and three hypermultiplets, with field content
\begin{eqnarray}\label{tab:BH}
&&\vbox{\hbox{\strut Betti-hyper truncation}\hrule}\nonumber\\
\mbox{gravity $+$ vector:}&\qquad&(g_{\mu\nu};A_1,k_{11}+k_{22};u_3)\,,\nonumber\\
\mbox{3 hypers:}&\qquad&(u_1,k,e_0^i,\tau,\bar\tau,b_0^i,\bar b_0^i,v,\bar v)\,.
\end{eqnarray}
The second truncation retains the $\mathcal N=2$ Betti vector multiplet, and yields two
vector multiplets and two hypermultiplets
\begin{eqnarray}\label{tab:BV}
&&\vbox{\hbox{\strut Betti-vector truncation}\hrule}\nonumber\\
\mbox{gravity $+$ 2 vectors:}&\qquad&(g_{\mu\nu};A_1,k_{11},k_{12};u_2,u_3)\,,\nonumber\\
\mbox{2 hypers:}&\qquad&(u_1,k,\tau,\bar\tau,b_0^i,\bar b_0^i)\,.
\end{eqnarray}
We will examine both of these truncations below.

Of course, it is possible to further truncate
away the entire Betti multiplet, leaving the universal $\mathcal N=2$ Sasaki-Einstein system
\begin{eqnarray}
&&\vbox{\hbox{\strut Sasaki-Einstein truncation}\hrule}\nonumber\\
\mbox{gravity $+$ vector:}&\qquad&(g_{\mu\nu};A_1,k_{11}+k_{12};u_3)\,,\nonumber\\
\mbox{2 hypers:}&\qquad&(u_1,k,\tau,\bar\tau,b_0^i,\bar b_0^i)\,.
\end{eqnarray}
If desired, the universal hypermultiplet may be truncated away, leaving
\begin{eqnarray}
&&\vbox{\hbox{\strut Massive vector truncation}\hrule}\nonumber\\
\mbox{gravity $+$ vector:}&\qquad&(g_{\mu\nu};A_1,k_{11}+k_{12};u_3)\,,\nonumber\\
\mbox{2 hypers:}&\qquad&(u_1,k,b_0^{m^2=21},\bar b_0^{m^2=21})\,.
\end{eqnarray}
Alternatively, we may also keep only the universal hypermultiplet
\begin{eqnarray}
&&\vbox{\hbox{\strut Universal hyper truncation}\hrule}\nonumber\\
\mbox{gravity:}&\qquad&(g_{\mu\nu};A_1+\ft13(k_{11}+k_{12}))\,,\nonumber\\
\mbox{hyper:}&\qquad&(\tau,\bar\tau,b_0^{m^2=-3},\bar b_0^{m^2=-3})\,.
\end{eqnarray}
Finally, all matter may be removed, leaving pure $\mathcal N=2$ supergravity
\begin{eqnarray}
&&\vbox{\hbox{\strut Pure sugra truncation}\hrule}\nonumber\\
\mbox{gravity:}&\qquad&(g_{\mu\nu};A_1+\ft13(k_{11}+k_{12}))\,.
\end{eqnarray}

In addition to the above family of truncations, it is possible to truncate IIB supergravity to
the NSNS sector before reducing.  Equivalently, we keep only fields arising from
$(g_{MN},\phi,F_3)$, where we have considered an S-duality rotated basis for convenience
in relating our results to the conifold.  The resulting NS truncation retains two vector
multiplets and two hypermultiplets
\begin{eqnarray}\label{tab:NS}
&&\vbox{\hbox{\strut NS truncation}\hrule}\nonumber\\
\mbox{gravity $+$ 2 vectors:}&\qquad&(g_{\mu\nu};A_1,b_1^2,b_2^2;\phi+4u_1,u_3)\,,\nonumber\\
\mbox{2 hypers:}&\qquad&(\phi-4u_1,u_2,c_0^2,e_0^2,b_0^2,\bar b_0^2,v,\bar v)\,.
\end{eqnarray}
As we show below, this is distinct from the Betti-vector truncation, even though they both
result in two vector multiplets and two hypermultiplets.  The NS truncation is related
to the baryonic branch of Klebanov-Strassler through a TST transformation
\cite{Maldacena:2009mw}.

In the following sub-sections we present the details of the
Betti-hyper, Betti-vector and the NS truncations. The theories are determined by the geometry of the special K$\ddot{a}$hler and quaternionic scalar coset manifolds. Along with some background information on the truncations, we provide only the particular Killing vectors which are gauged in each model as well as the form of the prepotentials. This is the most relevant information necessary to construct the superpotential and discuss the BPS flow equations.   Additional information for
each truncation will be relegated to the appendices.  For completeness, we present the reduction
of the IIB fermion supersymmetry variations in the appendices as well.  As a consistency check we
have verified that the Killing vectors and prepotentials determined from the coset and the fermion
reductions are in agreement.

\subsection{Betti-hyper truncation} \label{sec:bh}

We first consider the Betti-hyper truncation, which includes what is known as the
Betti-hypermultiplet \cite{Cassani:2010na}.  In total, it contains three $\mathcal N=2$
hypermultiplets and one vector multiplet. This field space has a critical point corresponding to the
Klebanov-Strassler solution and thus this truncation is of particular
interest. The supergravity theory is known to admit a superpotential
\cite{Liu:2011dw,Papadopoulos:2000gj}, but as we will discuss, this is not in fact a genuine superpotential but rather a {\it fake} superpotential.

The field content of the Betti-Hyper truncation is obtained from the
$\N=4$ theory by restricting to the modes which are invariant under
the $\cI$ symmetry:
\be
\cI=\Om_p\cdot (-1)^{F_L} \cdot \sig\,,
\ee
where
\bea
\Om_p\cdot (-1)^{F_L}:&&(g,\phi,B_{(2)},C_{(0)},C_{(2)},C_{(4)}) \ra  (g,\phi,-B_{(2)},C_{(0)},-C_{(2)},C_{(4)}) \,, \nonumber\\
\sig:&&(J_+,,J_-,\ROm,\IOm)\ra (J_+,,-J_-,-\ROm,-\IOm) \,.
\eea
The surviving field content is given in (\ref{tab:BH}), and additional details of the truncation are
presented in Appendix~\ref{app:BettiHyper}.

\subsubsection{Killing vectors}

The Killing vectors, which can be read off from the covariant derivatives in Section \ref{sec:N4}
 or from the supersymmetry variations in Appendix \ref{app:BettiHyper}, are
\begin{eqnarray}
K_0&=&-(Q+\epsilon_{ij}j_0^ie_0^j)\partial_k
-(3ib_0^i\partial_{b_0^i}+c.c)
+(\ft3{2}(1+\rho^2)\partial_\rho+c.c.)
-j_0^i\partial_{e_0^i} \,,
\nonumber\\
K_1&=&4\partial_k \,.
\end{eqnarray}

The corresponding Killing prepotentials can be obtained either from
the gravitino variation \eq{KSgr1} or by explicitly constructing the
$SU(2)$-connection $\om^r$ and the triplet of two-forms $\Om^r$ on the
hypermultiplet moduli space and then using \eq{PKOm}. In principle
these two methods should only agree up to a local $SU(2)$
transformation, but in fact we found them to agree precisely:
\begin{eqnarray}\label{eq:BHP}
P_{0}&=& -i[(\ft3{2\rho_2}(1+|\rho|^2) - \ft12e^{-4u_1}e^Z)\sigma_3
\nonumber\\
&&\kern1.5em-\ft{i}{2\rho}e^{-2u_1} v_i((\bar\rho-i)^2\bar f_0^i-(\bar\rho+i)^2f_0^i
+i(1-i\bar\rho)(1+i\bar\rho)j_0^i)\sigma_+\nonumber\\
&&\kern1.5em+\ft{i}{2\rho}e^{-2u_1}\bar v_i((\rho+i)^2f_0^i-(\rho-i)^2\bar f_0^i
-i(1+i\rho)(1-i\rho)j_0^i)\sigma_-]\,,
\nonumber\\
P_{1}&=& -2ie^{-4u_1}\sigma_3 \,.
\end{eqnarray}
Note that $P_{I}^{+}=\ol{(P_{I}^-)}$.

\subsubsection{The superpotential}

Much of the motivation of the current work is to understand the origin
in gauged supergravity of the superpotential first written down in
\cite{Papadopoulos:2000gj}:

\bea\label{eq:Wfake}
W_{KS} &=& - \fft12e^{-4u_1+4u_3}e^Z + 2e^{-4u_1-2u_3} + \fft3{2\rho}(1+|\rho|^2)e^{4u_3}\,.
\eea

Due to the particular form of the Killing prepotentials, namely that
$P_1^1=P_1^2=0$, the only non-trivial way to solve the algebraic prepotential constraint
\eq{crossvanish} is to set also $P_0^1=P_0^2=0$. This amounts to the condition
\be\label{eq:KScons}
3\fft{(1+i\bar\rho)}{(1-i\bar\rho)} v_ib_0^i + 3\fft{(1-i\bar\rho)}{(1+i\bar\rho)} v_i \bar b_0^i = v_i j_0^i \,.
\ee
Evaluated on this constraint, one finds
\be
\sqrt{P^rP^r}|_{\del_x Q^r=0} = P^3|_{\del_x Q^r=0}\,,
\ee
and thus the scalar potential can be obtained from the superpotential using the simple
potential from superpotential relation \eq{VWW}, so long as all quantities are subject to
the constraint $P_0^1=P_0^2=0$. What is particular interesting in this model is the
non-trivial fact that
\be
W_{KS}=P^3
\ee
recreates the scalar potential using \eq{VWW}, even without imposing any constraints. As a
result, $P^3$ plays the even more powerful role of a fake superpotential for this truncation.

In fact, nonsupersymmetric solutions of the KS system have been
studied in \cite{Kuperstein:2003yt}; in their analysis certain
solutions to the ``BPS" equations from the superpotential were shown
to correspond to $(3,0)$ flux on the deformed conifold, which is known
to be non-supersymmetric. From our analysis we can directly check that
these non-supersymmetric solutions do not satisfy the constraint
(\ref{eq:KScons}). Therefore they do not satisfy the true BPS
equations and are explicitly non-supersymmetric. We will elaborate on this point in
Section \ref{sec:WDCfake}.

\subsection{Betti-vector truncation}

We now turn to the Betti-vector truncation. Compared to the universal Sasaki-Einstein
truncation, this keeps an additional $\mathcal N=2$ vector multiplet as opposed to the
additional hypermultiplet of the Betti-hyper truncation, for a total of two hypermultiplets and two
vector multiplets. Details of this truncation are given in Appendix \ref{app:BettiVector}.  In
particular, we have the following three Killing vectors
\bea
K_0 &=& - (3ib_0^i\partial_{b_0^i} + c.c.) - Q\partial_k\,,  \non \\
K_1 &=&  2\partial_k \,, \non \\
K_2 &=&  2\partial_k\,,
\eea
and the prepotentials
\bea
P_{0} &=& -i\left[(3-\ft12e^{-4u_1}e^Z)\sigma_3 - 2ie^{-2u_1}v_if_0^i\sigma_+ + 2ie^{-2u_1}\bar v_i\bar f_0^i\sigma_- \right]\,, \non \\
P_{1} &=& -i e^{-4u_1}\sigma_3 \,, \non \\
P_{2} &=& -i e^{-4u_1}\sigma_3 \,.
\eea

Similar to the Betti-hyper truncation, the prepotentials $(P_1,P_2)$ are particularly simple.  This
again appears to be the key to constructing a fake superpotential from the $P^3$ term. From
$P^r \equiv X^I P^r_I$, where $X^I$ are given by
\be
X^0 = e^{4u_3}, \qquad X^1 = e^{2u_2-2u_3}, \qquad X^2 = e^{-2u_2-2u_3}\,,
\ee
we find
\bea\label{eq:WfakeBV}
W_{BV} &=& -\fft1{2}e^{-4u_1+4u_3}e^Z + e^{-4u_1-2u_2-2u_3}+e^{-4u_1+2u_2-2u_3}+3e^{4u_3} \,.\non \eea

As in the Betti-hyper truncation this superpotential acts as a fake superpotential. However, to our knowledge, the solution space of this has not been analyzed.  Of course, the fake superpotential
must be supplemented with the prepotential constraint \eq{crossvanish}, which in this case
takes on the particularly simple form
\be
v_i f_0^i=0\,,
\ee
and which is equivalent to two real constraints.

\subsection{NS-sector truncation} \label{sec:NS}

We now consider the NS-sector truncation.  This particular truncation on $T^{1,1}$ has not been previously worked out explicitly. However, its consistency is obvious from ten dimensions. We set the RR axion, the five-form and, for simplicity, the NSNS-three form to zero\footnote{By S-duality this is
related to a setup where only the NSNS-fields are non-vanishing.}. The resulting field content is listed in (\ref{tab:NS}), and the details of the truncation are given in Appendix~\ref{app:NS}.

In \cite{Maldacena:2009mw} this sector was shown to be related via a
TST transformation to the baryonic branch of the Klebanov-Strassler
theory. In the following we determine a superpotential for this sector
which in essence is then a superpotential on the baryonic
branch. However, we note that a fake superpotential in this sector has
not been found.

\subsubsection{Killing Vectors}

Again, the killing vectors can be determined from either the covariant derivatives in
Section~\ref{sec:N4} or the fermion variations in Appendix~\ref{app:NS}. They are
\bea
K_0 &=& - (3ib_0\partial_{b_0} + c.c.) + (3iv\partial_v + c.c.) - P\partial_{e_0} \,, \non \\
K_1 &=&  2\partial_{c_0}\,, \non \\
K_2 &=& 0\,.
\eea

The prepotentials, which can be computed from these Killing vectors on the scalar manifold or simply read off from the gravitino variation (\ref{eq:NSgrav}), are
\bea
P_{0} &=& -i\Big[\bigl(3 - \ft1{2}e^{\phi/2-2u_1}(e^{-2u_2}((1+|v|^2)j_0^2 + 2ivf_0 - 2i\bar v \bar f_0) - e^{2u_2}P)\bigr)\sigma_3 \non \\
&& \phantom{\ft{i}{\sqrt{6}}\Bigl[ }- \bigl(3\bar v + 2ie^{\phi/2-2u_1}(f_0 - \ft{i}2\bar v P)\bigr)\sigma_+
 - \bigl(3 v - 2ie^{\phi/2-2u_1}(\bar f_0 + \ft{i}2 v P)\bigr)\sigma_- \Bigr] \,,\non \\
P_{1}&=& -i\Big[e^{\phi/2-2u_1}(e^{-2u_2}(1-|v|^2)+e^{2u_2})\sigma_3 - 2\bar v e^{\phi/2-2u_1}\sigma_+ - 2 v e^{\phi/2-2u_1}\sigma_-\Big] \,,\non \\
P_{2} &=& 0\,,
\eea
where in the above, and for the remainder of this section, we have suppressed the upper
$SL(2,\mathbb R)$ index on the fields from the RR three-form and have set $j_0^2 = P$.

\subsubsection{The superpotential}

Curiously, we were not able to find a fake superpotential in this sector. This
seems to be related to the fact that $P_0$ and $P_1$ are both
non-trivial in all three components and so there is no natural $SU(2)$
direction for the prepotentials to lie.  This is in contrast to the previous two
truncations, which naturally fell into the 3-direction. One could argue that
these prepotentials can be rotated by an $SU(2)$ transformation into the
same form as in (\ref{eq:BHP}).
However, due to the nontrivial dependence of $P_1$ on the
hyper-scalars this rotation is field dependent and does not yield a
suitable fake superpotential. The key to constructing a fake
superpotential from prepotentials seems to be related to the fact
that theories which admit such a fake superpotential admit a {\it rigid} rotation of all non-trivial prepotentials into one direction. However, a rigorous demonstration of this statement has not been established.

Nevertheless, we may find the closest possibility for a superpotential in this sector by
computing $W = \sqrt{P^rP^r}$ and explicitly imposing the algebraic prepotential constraints
(\ref{crossvanish}) off-shell. In this case we find two independent constraints on the fields. The first is
\be
\Im (v b_0) = 0,
\ee
which can be solved by setting
\be
b_0 = \alpha \bar v,
\ee
where $\alpha$ is a real function. The second constraint is more
complicated and the detailed form is not illuminating. It however
fixes the coefficient $\alpha$ to be such that
\be\label{eq:NSconstraint}
b_0 = \fft{\left(2 P + 3e^{2u_1-2u_2-\phi/2}(1- |v|^2 - e^{4u_2})\right)}{6(1 + |v|^2 + e^{4u_2})}\bar v.
\ee
Once this identification has been made, the superpotential defined by
\be
W = \sqrt{\vec{P}\cdot\vec{P}},
\ee
can be used in the standard fashion and becomes
\bea\label{eq:NSW}
W &=& \sqrt{1+\ft14e^{-4u_2}(1-|v|^2-e^{4u_2})^2}  \non \\
&& \times
\left[2e^{-4u_1-2u_3}-Pe^{\phi/2-2u_1+4u_3}\left(\frac{1-|v|^2-e^{4u_2}}{1+|v|^2+e^{4u_2}}\right)
  + 6e^{4u_3-2u_2}\left(\frac{|v|^2+e^{4u_2}}{1+|v|^2+e^{4u_2}}\right)
  \right].\nonumber\\
\eea
It can be checked that once the constraint (\ref{eq:NSconstraint}) is
imposed, this expression for $W$ gives the potential, which is
also subject to  (\ref{eq:NSconstraint}), through the standard potential from
superpotential relation \eq{VWW}. A version of this superpotential, as well as the constraint (\ref{eq:NSconstraint}), has been previously derived in \cite{HoyosBadajoz:2008fw} in the context of a string dual to $\mathcal N = 1$ SQCD\footnote{We would like to thank I. Papadimitriou calling \cite{HoyosBadajoz:2008fw} to our attention.}. In \cite{HoyosBadajoz:2008fw}, Hamilton-Jacobi techniques are used to derive the superpotential in an effective one-dimensional scalar theory. This is somewhat different in philosophy to our analysis, where (\ref{eq:NSW}) is highlighted as a true superpotential within a genuine five-dimensional supergravity.

Note that the NS truncation includes the Maldacena-Nunez solution
\cite{Maldacena:2000yy}. In fact, substituting in the ansatz for the
IIB fields, the expression (\ref{eq:NSW}) reproduces the
superpotential shown in \cite{Papadopoulos:2000gj}. Moreover, we can
verify that the more generic ansatz of \cite{Maldacena:2009mw} obeys
the BPS flow equations derived from this superpotential. Therefore,
via the TST transformation detailed in \cite{Maldacena:2009mw}, this
superpotential in fact describes the baryonic branch of the
Klebanov-Strassler theory.

\section{Discussion}\label{sec:conc}

The coset reduction of IIB supergravity on $T^{1,1}$ naturally yields five-dimensional
gauged $\mathcal N=4$ supergravity.  We have analyzed three particular $\N=2$ truncations
of this reduction that are relevant to the conifold solution and its relatives.  In particular,
we have highlighted the difference between fake and real superpotentials and demonstrated
the importance of the prepotential constraint \eq{crossvanish} as a necessary condition for
the supersymmetry of the solutions.

\subsection{Fake superpotentials and the warped deformed conifold} \label{sec:WDCfake}

There is a particularly relevant class of solutions within the Betti-hyper truncation which correspond to taking the ten-dimensional IIB background to be a warped product of $\RR^{1,3}$ and the Ricci-flat
metric on the deformed conifold. We can solve this system explicitly using the fake superpotential (\ref{eq:Wfake}). In particular, this amounts to specifying the fields coming from the metric to take the form of the deformed conifold metric and solving the flow equations with the fake superpotential (\ref{eq:Wfake}). This is a particularly nice example to study in the context of fake superpotentials as there exists a known non-supersymmetric solution to the flow equations derived from (\ref{eq:Wfake}), found in \cite{Kuperstein:2003yt}.

In order to make the connection with previous solutions as transparent
as possible we define the flux of the NS and RR three forms to be
$j_0^1 = R$ and  $j_0^2 = P$, respectively, and make the following
KS-like parametrization for the other scalars in the three forms:
\bea\label{eq:KSlike}
b_0^1 = -\frac{R}3(\tilde{F} - \fft12) - i \frac{P}6(f_{KS} - k_{KS}), && e_0^1 = \frac{P}3(f_{KS} + k_{KS}), \nonumber\\
b_0^2 = -\frac{P}3(F_{KS} - \fft12) + i \frac{R}6(\tf - \tk), && e_0^2 = -\frac{R}3(\tf + \tk).
\eea
The functions $f_{KS}$, $k_{KS}$, and $F_{KS}$ are the standard
functions in the KS ansatz, and the tilde-ed functions $\tf$, $\tk,$ and
$\tilde{F}$ are their S-dual analogs. Assuming a vanishing
axion, $a = 0,$ the equations reduce to two decoupled systems for
$\{f_{KS},k_{KS},F_{KS}\}$ and $\{\tf,\tk,\tilde{F}\}$ and the
solution is given by \cite{Kuperstein:2003yt}:
\bea
f_{KS}(t)&=& \frac{(-t \coth{t} + 1)}{2\sinh{t}}(-1+\cosh{t})\nonumber \\
&& +  C_1\bigg(-t +\fft12\sinh{t} + \frac{t}{2(1+\cosh{t})} +
\fft12\tanh{\ft{t}2}\bigg) - \frac{C_2}{1+\cosh{t}} + C_3, \nonumber\\
k_{KS}(t)&=& \frac{(-t \coth{t} + 1)}{2\sinh{t}}(1+\cosh{t}) \nonumber\\
&&  + C_1\bigg(-t -\fft12\sinh{t} - \frac{t}{2(-1+\cosh{t})} +
\fft12\coth{\ft{t}2}\bigg)  - \frac{C_2}{1-\cosh{t}} + C_3, \nonumber\\
F_{KS}(t) &=& \fft12 - \frac{t}{2\sinh{t}} + \fft12C_1\left(\cosh{t} -
\frac{t}{\sinh{t}}\right) + \frac{C_2}{\sinh{t}},
\eea
where $C_1$, $C_2,$ and $C_3$ are integration constants. Additionally,
the solution for the tilde-ed functions is exactly the same, but with
different integration constants  $\tC_1$, $\tC_2,$ and $\tC_3$.

The solution to the ``KS" system ({\it i.e.}\ with $R=0$) has already been solved in \cite{Kuperstein:2003yt}, yielding the above solution. The only
non-singular solution in this sector is with $C_1 = C_2 = C_3 = 0$
which reduces exactly to the Klebanov-Strassler solution. In
\cite{Kuperstein:2003yt}, it was also noted
that the solution with $C_1 = 1$ and $C_2 = C_3 = 0$ corresponds to a
background with $(0,3)$-flux which breaks supersymmetry by arguments from string theory \cite{Grana:2001xn}. In the
present context we can verify explicitly that this solution is not supersymmetric by evaluating the two constraints $P_0^1 = 0$ and $P_0^2 = 0$. The explicit
form of the constraints is not important. However we find that $P_0^1
 \propto \tC_1$ and $P_0^2 \propto C_1.$ This means that solutions
with $C_1$ or $\tC_1$ non-vanishing are not supersymmetric. In particular, we see that the non-supersymmetric solution found in \cite{Kuperstein:2003yt} is due to the superpotential
\eq{eq:Wfake} being a fake superpotential.  In this case, solving the first order flow equations is
insufficient in itself in guaranteeing supersymmetry, and the algebraic prepotential conditions
must also be checked.

In fact, there is a subtlety in obtaining non-supersymmetric solutions using the fake
superpotential.  Ordinarily, solving the first order BPS equations will ensure a solution to the
bosonic equations of motion.  However, if the prepotential conditions are not satisfied, there is
at least a possibility that the system may not solve the full set of equations of motion.  In the
present case, there would be a concern that the fluxes $j_0^1 = R$ and $j_0^2 = P$ along with
non-trivial scalar profiles for $e_0^i$ as well as the complex charged scalars $b_0^i$ may source
the graviphoton $A_1$.  However, we have checked that the source for $A_1$ vanishes
regardless of the choice of integration constants $C_i$ and $\tC_i$.  Hence the solution is
valid in both the supersymmetric and non-supersymmetric cases.

Note that since both $F_3$ and $H_3$ are nonzero, the five-form is sourced so that in addition to the flux term in the original KS solution, which
is encoded in $e^Z$, the scalar $k$ is, in general, non-zero as
well. The explicit form of $k$ is not so illuminating. However it vanishes for
the non-singular solution when all integration constants are set to zero.

The notion  of non-supersymmetric flux on warped Calabi-Yau backgrounds has been generalized in \cite{Lust:2008zd} to include $SU(3)\times SU(3)$ structure backgrounds. It would be interesting to connect those ideas to the existence of a fake superpotential in five dimensions for some more general truncation than those considered in this work.

\subsection{Superpotential for the baryonic branch of the warped deformed conifold}

One distinguishing feature of the baryonic branch of the warped deformed conifold is that away from the origin it breaks the $\ZZ_2$ symmetry which we call $\cI$. The NS truncation we considered includes $\ZZ_2$ odd and even modes and within this theory there is a line of half-BPS solutions \cite{Butti:2004pk}. A very neat observation of \cite{Maldacena:2009mw} is that one can perform a certain TST transformation on this family of solutions and connect it to the family which is dual to the baryonic branch of the warped deformed conifold. Physically this latter solution space is more interesting since the whole family is dual to quantum field theory.

In principle it is possible to make a five dimensional domain wall ansatz and then perform the TST transformation on the full theory off-shell. This is quite an unwieldy operation, but it would interesting to work out a way to characterize this transformation covariantly in terms of the scalar cosets of the NS truncation.

One motivation for uncovering a superpotential for the baryonic branch is to study perturbation of the warped deformed conifold along the lines of \cite{Borokhov:2002fm, Bena:2009xk}. For those works the superpotential used only included $\ZZ_2$-even modes. But using the superpotential computed in this work, it should be possible to include $\ZZ_2$-odd modes in the NS sector and then use the TST transformation to map them to genuine perturbations of the warped deformed conifold.

\vspace{1cm}
\noindent {\bf \Large Acknowledgements:}
\vspace{1cm}

P.S. would like to thank Ibrahima Bah and Alberto Faraggi for useful discussions. This work was supported in part by the US Department of Energy under grants DE-FG02-95ER40899 and DE-FG02-97ER41027. The work of N.H. was support by NSF grant PHY-0804450 and by the grant ANR-07-CEXC-006.

\appendix

\section{Details of the Betti-hyper truncation}\label{app:BettiHyper}

Here we present some additional details of the Betti-hyper truncation.  This truncation gives
rise to $\mathcal N=2$ gauged supergravity coupled to one vector multiplet and three
hypermultiplets.  The bosonic fields in the gravity and vector multiplet are
$(g_{\mu\nu};A_1,k_{11}+k_{22};u_3)$, and the 12 scalars in the hypermultiplet are
$(u_1,k,e_0^i,\tau,\bar\tau,b_0^i,\bar b_0^i,v,\bar v)$.

\subsection{Bosonic sector}

The full Lagrangian is
\be
\cL=\cL_{gr}+\cL_{hyp} + \cL_{vec}+\cL_{g,kin}+\cL_{CS}+\cL_{pot},\label{FullLag}
\ee
where the individual components are given below.

\subsubsection{Hypermultiplet sector}

The hypermultplet kinetic terms are
\bea
\cL_{hyp}&=& -e^{-4u_1}\mathcal{M}_{ij}\Bslb \half e^{-4u_2}\hat g^i_{11}\w * \hat g^j_{11} + \half e^{4u_2} \hat g_{12}^i \w * \hat g_{12}^j + 2(\hat f^i_1\w *\hat{\bar f}^j_1 + \hat{\bar f}^i_1\w *\hat f^j_1 ) \Bsrb \non \\
&&-8d u_1\w*d u_1  -4d u_2\w*d u_2 -12 d u_3\w*d u_3 - e^{-4u_2} (d|v| \w * d|v|+|v|^2D\tha \w *D \tha ) \non \\
&&-\half e^{-8u_1} K_1\w * K_1  -\half d\phi \w * d\phi -\half e^{2\phi} da \w *da \label{Lskin},
\eea
with the relation
\be
e^{2u_2}=\frac{1}{\cosh y}\,, \ \ |v|=\tanh y \,,
\ee
where
\bea
\hat g^i_{11}&=& De_0,  \non\\
\hat g^i_{12}&=& (1+|v|^2)De^i_0 - 4 \Im \!(vDb^i_0), \non \\
\hat f_1^i &=& Db_0^i - \fft{i}2\vbar De_0^i,\non \\
De^i_0 &=& de^i_0 - j_0^i A_1, \non \\
Db^i_0 &=& db^i_0 - 3ib_0^i A_1, \non \\
D\tha&=& d\tha+3 A_1,
\eea
and
\be
\mathcal M=e^{\phi}\begin{pmatrix}a^2+e^{-2\phi}&-a\cr-a&1\end{pmatrix}.
\ee

Following \cite{Lu:1998xt}, the generators of the solvable subalgebra of $SO(4,3)$ may be
taken as
\bea
H_1= e_{11}-e_{55}\,, && H_2= e_{22}-e_{66}\,,\ \ H_3= e_{33}-e_{77}\,,\,,  \non \\
E_1^{\ 2}= -e_{21}+e_{56} \,, &&E_1^{\ 3}= -e_{31}+e_{57}\,,\ \ E_2^{\ 3}= -e_{32}+e_{67} \non\\
V^{12}= e_{16}-e_{25}\,, &&V^{13}= e_{17}-e_{35}\,,\ \ V^{23}= e_{27}-e_{36}\,, \non \\
U_{1}^{1}= e_{14}+e_{45} \,, &&
U_{1}^{2}= e_{24}+e_{46}\,, \ \
U_{1}^{3}= e_{34}+e_{47}\,.
\eea
Using these, the metric on the hyperscalar coset is
\bea
-\frac{1}{8}\Tr dM \w *dM^{-1}&=&\frac{1}{4} \blp d\phi_1^2+d\phi_2^2+d\phi_3^2\brp \non \\
&&\hspace{-2cm}+ \half e^{-\phi_1+\phi_2} dx_4^2 + \half e^{\phi_3-\phi_1}\blp dx_{5}+x_4dx_6\brp^2+\half e^{-\phi_2+\phi_3} dx_6^2 \non \\
&& \hspace{-2cm}+\half e^{\phi_1+\phi_2}\blp  dx_7 +x_{10}dx_{11}  \brp^2
+\half e^{\phi_1+\phi_3} \blp dx_8 -x_6 dx_7+x_{10}(dx_{12}-x_6 dx_{11}) \brp^2 \non \\
&&\hspace{-2cm} +\half e^{\phi_2+\phi_3} \blp dx_9 -x_4 dx_8 +(x_5+x_4x_6)(dx_7+x_{10}dx_{11})+(x_{11}-x_4x_{10})dx_{12} \brp^2\non \\
&&\hspace{-2cm}+\half e^{\phi_1}dx_{10}^2
+\half e^{\phi_2}\blp dx_{11}-x_4 dx_{10} \brp^2
+\half e^{\phi_3}\blp dx_{12}-x_5 dx_{10}-x_6dx_{11} \brp^2  \non\,, \\
\eea
where
\be
M=L^T L,
\ee
and
\be
L=e^{\frac{\phi_1}{2} H_1}\cdot e^{\frac{\phi_2}{2} H_2}\cdot e^{\frac{\phi_3}{2} H_3}\cdot
e^{x_4 E_{1}^{\ 2}}\cdot e^{x_5 E_{1}^{\ 3}}\cdot e^{x_6 E_{2}^{\ 3}}\cdot e^{x_7 V_{12}}\cdot
e^{x_8 V_{13}}\cdot e^{x_9 V_{23}}\cdot e^{x_{10} U_{1}^1}\cdot e^{x_{11} U_{2}^1}\cdot
e^{x_ {12} U_{3}^{1}}\,.
\ee

The supergravity fields and the coset fields are related by the coordinate transformations
\bea
 \phi_1  &=&2x-2\pi i\,,\non \\
\phi_2&=& -4u_1-\phi\,, \non \\
\phi_3&=&-4u_1+\phi\,, \non \\
 x_4&=& e_0^1 + 2b_{0i}^1\,, \non \\
x_5  &=& e_0^2 - a e_0^1 + 2b_{0i}^2\,, \non \\
x_6&=&a\,, \non \\
 x_{7}&=&e_0^1(1-\chi^2)- 2b^1_{0i} (1+\chi^2) \,, \non\\
x_{8}&=& e_0^2(1-\chi^2)-2b^2_{0i}(1+\chi^2) \,,\non \\
x_{9}&=& k - 4b^1_{0r}b^2_{0r} -2(e_0^1b^2_{0i} - e_0^2b^1_{0i}) + 4\chi b^1_{0r}(e_0^2+2b^2_{0i}) - \chi^2(e_0^1+2b^1_{0i})(e_0^2+2b^2_{0i})\,,\non \\
x_{10} &=& \sqrt{2}\chi\,,  \non\\
 x_{11}&=& \sqrt{2}\blp -2b^1_{0r} + \chi(e_0^1+2b^1_{0i})\brp \,. \non\\
x_{12}&=&\sqrt{2}\blp -2b^2_{0r} + \chi(e_0^2+2b^2_{0i})\brp \,.
\eea
In these coordinates, we note that
\be
\rho=\chi+i e^{-x}
\ee
is an $SL(2,\RR)$ factor within the coset which descends from the scalar $v$ by the identification
\be
v = -\frac{(i-\rho)(i-\bar\rho)}{1+|\rho|^2}.
\ee

\subsubsection{Vector multiplet sector}

The vector multiplet kinetic terms and Chern-Simons terms are
\bea
\cL_{vec}&=&-12 du_3\w  * du_3, \\
\cL_{g,kin}&=&-\half e^{-8u_3} F_2\w * F_2-e^{4u_3} K_{2}\w * K_{2}, \\
\cL_{CS}&=&- A\w K_2 \w K_2.
\eea
This and the supersymmetry variations lead to the identification of the constrained scalars as
\be
X^0 = e^{4u_3}, \qquad X^1 = e^{-2u_3},
\ee
with $c_{011} = 2$. The field strengths are given by%
\footnote{Note that the subscripts on $F_2$ and $K_2$ refer only to the degree of the forms
and are not special geometry indices.}
\be
F^0 = F_2 , \qquad F^1 = -K_2.
\ee

\subsubsection{Scalar potential}

The scalar potential has several contributions which we distinguish for clarity:
\bea
\cL_{pot}&=&-\blp V_{gr}+V_{F_{(3)}}+V_{F_{(5)}}\brp, \label{potential}\\
V_{gr}&=&  -12e^{-4u_1-2u_2+2u_3}\blp 1+|v|^2 +e^{4u_2}\brp  +9|v|^2e^{-4u_2+8u_3}  \non \\
&&+2e^{-8u_1-4u_3} \blp e^{4u_2}+e^{-4u_2}(1-|v|^2)^2+2|v|^2\brp, \\
V_{F_{(3)}}&=&\fft12 e^{-4u_1+8u_3}\mathcal{M}_{ij}\Blp e^{-4u_2} \hat j^i_{01}\hat j^j_{01} +  e^{4u_2} \hat j^i_{02}\hat j^j_{02} + 2(\hat f^i_0 \hat{\bar f}^j_0 + \hat{\bar f}^i_0 \hat f^j_0)\Brp, \\
V_{F_{(5)}}&=& \half e^{2Z} e^{-8u_1+8u_3}\,,
\eea
where
\bea
e^{4u_2} &=& 1- |v|^2, \non \\
\hat j^i_{01} &=& (1+|v|^2) j^i_0 - 4\Im \!(f^i_0v), \non \\
\hat j^i_{02} &=& - j_0^i, \non \\
\hat f_0^i &=& f^i_0 - \fft{i}2j^i_0 \vbar, \non \\
e^Z &=& Q - \fft{2i}3\epsilon_{ij}(f_0^i\bar f_0^j-\bar f_0^if_0^j) + \epsilon_{ij}(j_0^i e_0^j - j_0^j e_0^i).
\eea

\subsection{Fermion variations}

The supersymmetry variations of the KS-sector have been worked out in \cite{Liu:2011dw}, where the fermions were organized according to mass eigenstates of the fluctuations on the AdS$_5$ background solution.  However, in terms of $\mathcal N = 2$ gauged supergravity, they are more naturally organized into variations appropriate for three hypermultiplets and one vector multiplet. This is accomplished by defining the following linear combinations of the AdS$_5$ mass eigenstates as the three hyperini and one gaugino
\bea
\zeta^1 &=& -\lambda^c, \non\\
\zeta^2 &=& -\left(\frac{1+|\rho|^2}{1+\bar\rho^2}\right)\psi^{m=-3/2}, \non \\
\zeta^3 &=& -\ft1{15}(2\psi^{m=11/2}-3\psi^{m=-9/2}), \non \\
\xi^1 &=& \ft1{5}(\psi^{m=11/2}+\psi^{m=-9/2}),
\eea
where $\rho = \chi + ie^{-x}$ is the $SL(2,\mathbb R)$ scalar descending from $v$.

Furthermore we define a phase rotated supersymmetry parameter $\varepsilon'$ as
\be
\varepsilon = \left(\frac{\rho+i}{\bar\rho-i}\right)^{1/2} \varepsilon'.
\ee
We similarly rotate the $\zeta^i$, $\xi^1$ and the gravitino,
\bea
\zeta^i &=& \left(\frac{\rho+i}{\bar\rho-i}\right)^{1/2}\zeta^{i'},\non \\
\xi^1 &=& \left(\frac{\rho+i}{\bar\rho-i}\right)^{1/2}\xi^{1'}, \non \\
\psi_\alpha &=& \left(\frac{\rho+i}{\bar\rho-i}\right)^{1/2}\psi_\alpha'.
\eea
With these identifications, the supersymmetry transformations are
\bea
\delta\zeta^{1'} &=& \left(-\ft{i}2\gamma\cdot\partial\phi - \ft12e^\phi\gamma\cdot\partial a\right)\varepsilon' + \fft{ie^{-2u_1}}{4\tau_2} \bar v_i \Big[i(1+\bar\rho^2)\big(\gamma\cdot De_0^i - ie^{4u_3}j_0^i\big) \non \\
&& + (\bar\rho-i)^2\big(\gamma\cdot f_1^i -ie^{4u_3}f_0^i\big)- (\bar\rho+i)^2\big(\gamma\cdot\bar f_1^i-ie^{4u_3}\bar f_0^i\big) \Big](\varepsilon')^c, \non\\
\delta\zeta^{2'} &=& \left(\ft{1}{2\rho_2}\gamma\cdot D\rho + \ft{3i}{2\rho_2}e^{4u_3}(1+\rho^2)\right)\varepsilon' - \fft1{2}e^{-2u_1} v_i \Big[\big(\gamma\cdot De_0^i - ie^{4u_3}j_0^i\big) \non \\
&& - i\frac{(\rho-i)(\bar\rho-i)}{1+|\rho|^2}\big(\gamma\cdot f_1^i -ie^{4u_3}f_0^i\big)+ i\frac{(\rho+i)(\bar\rho+i)}{1+|\rho|^2}\big(\gamma\cdot\bar f_1^i-ie^{4u_3}\bar f_0^i\big) \Big](\varepsilon')^c,\non\\
\delta\zeta^{3'} &=& \left[-\ft{i}2\gamma\cdot\partial u_1 - \ft18e^{-4u_1}\gamma\cdot K_1 - \ft{i}2e^{-4u_1-2u_3}+\ft{i}8e^{-4u_1+4u_3}e^Z\right]\varepsilon' \non \\
&& - \fft{e^{-2u_1}}{16\tau_2} v_i \Big[(\bar\rho-i)^2\big(i\gamma\cdot f_1^i - e^{4u_3}f_0^i\big) - (\bar\rho+i)^2\big(i\gamma\cdot \bar f_1^i - e^{4u_3}\bar f_0^i\big)  \non \\
&& \phantom{- \fft{e^{-2u_1}}{16\rho_2} v_i \Big[} - (1+\bar\rho^2)\big(\gamma\cdot De_0^i + ie^{4u_3}j_0^i\big)\Big](\varepsilon')^c , \non\\
\delta\xi^{1'} &=& \Bigl[\ft{i}2\gamma\cdot\partial u_3 + \ft1{24}e^{-4u_3}\gamma\cdot(F_2+e^{6u_3}K_2) + \ft{i}6e^{-4u_1-2u_3} - \ft{i}{4\rho_2}(1+|\rho|^2)e^{4u_3} \nonumber\\
&&+ \ft{i}{12}e^{-4u_1+4u_3}e^Z\Bigr]\varepsilon'  + \fft{e^{-2u_1+4u_3}}{12\rho_2} v_i \Big[(\bar\rho-i)^2f_0^i - (\bar\rho+i)^2\bar f_0^i + i(1+\bar\rho^2)j_0^i \Big](\varepsilon')^c,\nonumber\\
\eea
where $D\rho \equiv d\rho - \ft32 (1+\rho^2)A_1$.
The gravitino variation is
\bea
\delta\psi_\alpha' &=& \Big[\mathcal D_\alpha + \ft{i}{24}(\gamma_\alpha{}^{\beta\gamma} - 4\delta\alpha^\beta\gamma^\gamma)(e^{-4u_3}F_{\beta\gamma}-2e^{2u_3}K_{2\,\beta\gamma}) \non \\
&& + \ft16 \gamma_\alpha\left(2e^{-4u_1-2u_3} + \ft3{2\tau_2}(1+|\rho|^2)e^{4u_3} - \ft12e^{-4u_1+4u_3}e^Z\right)\Big]\varepsilon' \non \\
&& -\ft{i}{12\rho_2}e^{-2u_1+4u_3}v_i\gamma_\alpha\left[(\bar\rho-i)^2f_0^i - (\bar\rho+i)^2\bar f_0^i + i(1+\bar\rho^2)j_0^i\right](\varepsilon')^c, \label{KSgr1}
\eea
where the supercovariant derivative acts as
\bea
\mathcal D_\alpha \varepsilon' &=& \left(\nabla_\alpha -\ft{3i}2A_\alpha - \ft{i}4e^{-4u_1}K_{1\,\alpha} + \ft{i}4e^{\phi}\partial_\alpha a - \ft{i}{2\rho_2}\partial_\alpha\rho_1\right)\varepsilon'  \non \\
&& + \ft1{4\rho_2}e^{-2u_1}v_i\Big((\bar\rho-i)^2f_{1\,\alpha}^i - (\bar\rho+i)^2\bar f_{1\,\alpha}^i + i(1+\bar\rho^2)D_\alpha e_0^i\Big)(\varepsilon')^c.
\eea
Here we have written the terms from the three-forms using the $SL(2,\mathbb R)$ vielbein $v_i$ where $
v_1 = -(ae^{\phi/2} + i e^{-\phi/2})$ and $v_2 = e^{\phi/2},$ such that the complex three-form takes the form
\be
\fft1{\sqrt{\tau_2}}G_3 = v_iF_3^i.
\ee

\section{Details of the Betti-vector truncation} \label{app:BettiVector}

The Betti-vector truncation yields $\mathcal N=2$ gauged supergravity coupled to two
vector multiples and two hypermultiplets.  The bosonic fields in the gravity and vector multiplets
are $(g_{\mu\nu};A_1,k_{11},k_{12};u_2,u_3)$, and the eight scalars in the hypermultiplet are
$(u_1,k,\tau,\bar\tau,b_0^i,\bar b_0^i)$.

\subsection{Bosonic sector}

The full Lagrangian is
\be
\cL=\cL_{gr}+\cL_{hyp} + \cL_{vec}+\cL_{g,kin}+\cL_{CS}+\cL_{pot},
\ee
where the individual components are given below.

\subsubsection{Hypermultiplet sector}

The hypermultplet kinetic terms are
\bea
\cL_{hyp}&=&- 4 e^{-4u_1+\phi} \mathcal{M}_{ij}f^i_1 \w * \bar{f}^j_1 - 8d u_1\w*d u_1  -4d u_2\w*d u_2 -12 d u_3\w*d u_3  \non \\
&&-\half e^{-8u_1} K_1\w * K_1  -\half d\phi \w * d\phi -\half e^{2\phi} da \w *da\, ,
\eea
where
\bea
f_1^i &=& Db^i_0, \non \\
K_1 &=& Dk + 2\epsilon_{ij}[b_0^i D\bar{b}_0^j + \bar b_0^i Db_0^j], \non \\
Dk &=& dk - Q A_1 - 2k_{11} - 2k_{12}.
\eea

Using the conventions of \cite{Lu:1998xt}, the metric on the coset
\be
\cM_{\rm hyp}=\frac{SO(4,2)}{SO(4)\times SO(2)} \label{hyperman}
\ee
is
\bea
-\frac{1}{8}\Tr dM \w *dM^{-1}&=&\frac{1}{4} \blp d\phi_1^2+d\phi_2^2\brp + \half e^{-\phi_1+\phi_2} dx_1^2 + \half e^{\phi_1}\blp dx_3^2+dx_4^2\brp \non\\
&&+ \half e^{\phi_2} \blp  (d(x_5-x_1x_4)+x_4dx_1)^2+(d(x_6-x_1x_3)+x_3dx_1)^2\brp \non \\
&& + \half e^{\phi_1+\phi_2} \blp dx_2+x_3dx_6+x_4dx_5\brp^2,
\eea
which is related to  $\cL_{hyp}$ by the field redefinitions
\bea
-\phi-4u_1&=&\phi_1,\non \\
\phi-4u_1 &=& \phi_2,\non \\
a &=& x_1,\non \\
k&=& x_2+\half  x_1(x_3x_6+x_4x_5),\non\\
2\sqrt{2} b_0^1 &=& x_4 - ix_3,\non \\
2\sqrt{2} b_0^2 &=&  x_5-x_1 x_4 - i(x_6-x_1x_3)\,.
\eea

\subsubsection{Vector multiplet sector}

The scalars in the vector multiplets have
\bea
\cL_{\rm vec}&=& -12 du_3\w *du_3 -\frac{1}{4}d (4u_1+\phi)\w*d (4u_1+\phi). \label{svec}
\eea
The gauge kinetic terms are
\bea
\cL_{g,kin}&=&-\half e^{-8u_3} F_2\w * F_2 -\half e^{-4u_2+4 u_3} K_{21}\w * K_{21}- \half e^{4u_2+4u_3} K_{22}\w *K_{22}.  \label{Lgkin}
\eea
There is also the Chern-Simons coupling
\be
\cL_{CS}=  -A_1 \w K_{21} \w K_{22}.
\ee

From $\cL_{vec}$ we see that the two real scalars in the vector multiplets are $u_3$ and $u_2$ and they parameterize the manifold
\be
\cM_v=SO(1,1)\times SO(1,1).
\ee
The special geometry data for this case is given by the constrained scalars
\be
X^0 = e^{4u_3}, \qquad X^1 = e^{2u_2-2u_3}, \qquad X^2 = e^{-2u_2-2u_3},
\ee
with $c_{012} = 1$, and the vector field strengths are given by
\be
F^0 = F_2, \qquad F^1 = -K_{21}, \qquad F^2 = K_{22}.
\ee

\subsubsection{Scalar potential}

The scalar potential has several contributions which we distinguish for clarity:
\bea
\cL_{pot}&=&-\blp V_{gr}+V_{F_{(3)} }+ V_{F_{(5)}}\brp, \label{potentialBV}\\
V_{gr}&=&  -12e^{-4u_1-2u_2+2u_3}\blp 1 +e^{4u_2}\brp +2e^{-8u_1-4u_3} \blp e^{4u_2}+e^{-4u_2}\brp, \\
V_{F_{(3)}}&=& 2 e^{-4u_1+8u_3} \mathcal{M}_{ij}(f^i_0\fbar^j_0 + \fbar^i_0f^j_0) ,   \\
V_{F_{(5)}}&=& \half e^{2Z} e^{-8u_1+8u_3},
\eea
where
\be
e^Z = Q - \fft{2i}3\epsilon_{ij}(f_0^i\bar f_0^j-\bar f_0^if_0^j).\label{eZ}
\ee

\subsection{Fermion variations}

The supersymmetry variations were worked out in \cite{Liu:2011dw}. We again organize the fermions into linear combinations appropriate to the $\mathcal N = 2$ multiplet identifications as opposed to the mass eigenstates. In particular, we define
\bea
\zeta^1 &=& -\lambda^c, \non\\
\zeta^2 &=& \ft12(\psi^{m=11/2}+\psi^{m=-9/2}), \non \\
\xi^1 &=& -\psi^{m=-1/2}, \non \\
\xi^2 &=& -\ft1{15}(2\psi^{m=11/2}-3\psi^{m=-9/2}),
\eea
where $\zeta^i$ are the two hyperini and $\xi^i$ are the gaugini.
The supersymmetry transformations are then
\bea
\delta\zeta^{1} &=& \left(-\ft{i}2\gamma\cdot\partial\phi - \ft12e^\phi\gamma\cdot\partial a\right)\varepsilon - e^{-2u_1} \bar v_i \left(i\gamma\cdot f_1^i+e^{4u_3}f_0^i\right)\varepsilon^c, \non\\
\delta\zeta^2 &=& \left(-\ft{i}2\gamma\cdot\partial u_1 - \ft18e^{-4u_1}\gamma\cdot K_1 -\ft{i}4e^{-4u_1-2u_3}(e^{-2u_2}+e^{2u_2} -\ft12e^{6u_3}e^Z)\right)\varepsilon \non \\
&& - e^{-2u_1} v_i \left(i\gamma\cdot f_1^i + e^{4u_3}f_0^i\right)\varepsilon^c ,\non\\
\delta\xi^1 &=& \left(-\ft{i}2\gamma\cdot\partial u_2 + \ft1{16}e^{2u_3}\gamma\cdot(e^{-2u_2}K_{21} - e^{2u_2}K_{22}) - \ft{i}4e^{-4u_1-2u_3}(e^{-2u_2}-e^{2u_2})\right)\varepsilon, \non \\
\delta\xi^2 &=& \big(-\ft{i}2\gamma\cdot\partial u_3 + \ft1{24}e^{-4u_3}\gamma\cdot(F_2+\ft12e^{-2u_2+6u_3}K_{21} + \ft12e^{2u_2+6u_3}K_{22}) + \ft{i}{12}e^{-4u_1+4u_3}e^Z \non\\
 && + \ft{i}{12} ( e^{-4u_1-2u_2-2u_3} + e^{-4u_1+2u_2-2u_3} - 6e^{4u_3})\big)\varepsilon  - \ft13 e^{-2u_1+4u_3} v_i f_0^i \varepsilon^c.
\eea
Finally, the gravitino variation is
\bea
\delta\psi_\alpha
&=&\big(D_\alpha +\ft{i}{24}(\gamma_\alpha{}^{\beta\gamma}
-4\delta_\alpha^\beta\gamma^\gamma)(e^{-4u_3}F_{\beta\gamma} - e^{-2u_2+2u_3}K_{1\,\beta\gamma}
- e^{2u_2+2u_3}K_{2\,\beta\gamma}) \nonumber \\
&&+\ft1{6}\gamma_\alpha(e^{-4u_1-2u_2-2u_3}+e^{-4u_1+2u_2-2u_3}+3e^{4u_3}-\ft1{2}e^{-4u_1+4u_3}e^Z) \big)\varepsilon \non \\
&&-\ft{i}3\gamma_\alpha e^{-2u_1+4u_3}v_if^i_0 \varepsilon^c,
\eea
where the covariant derivative acts on the supersymmetry parameter as
\be
D_\alpha\varepsilon \equiv  \left(\nabla_\alpha - \ft{3i}2 A_\alpha
- \ft{i}4e^{-4u_1}K_{1\,\alpha} + \ft{i}{4}e^\phi\partial_\alpha a\right)\varepsilon + e^{-2u_1}v_if^i_\alpha\varepsilon^c.
\ee

\section{Details of the NS truncation}\label{app:NS}

The final $\mathcal N=2$ truncation is to the NS sector of IIB supergravity.  The resulting
truncation has two vector multiplets and two hypermulitplets.  The bosonic fields in the
gravity and vector multiplets are $(g_{\mu\nu};A_1,b_1^2,b_2^2;\phi+4u_1,u_3)$ and
the eight scalars in the hypermultiplet are
$(\phi-4u_1,u_2,c_0^2,e_0^2,b_0^2,\bar b_0^2,v,\bar v)$.

\subsection{Bosonic Sector}

The full Lagrangian is
\be
\cL=\cL_{gr}+\cL_{hyp}+\cL_{vec}+\cL_{g,kin}+\cL_{pot}\label{FullLagNS} \,.
\ee
The individual components are given below.

\subsubsection{Hypermultiplet Sector}

The hypermulitplet kinetic terms are
\bea
\cL_{hyp}&=&-\half e^{-4(u_1+u_2)+\phi} \hat g_{11}\w * \hat g_{11} -\half e^{-4(u_1-u_2)+\phi}\hat g_{12}\w * \hat g_{12}  - 4 e^{-4u_1+\phi}\hat f_1 \w * \hat{\bar{f}}_1   \non \\
&&-\frac{1}{4}d (4u_1-\phi)\w*d (4u_1-\phi)  -4d u_2\w*d u_2  - e^{-4u_2} Dv \w * D\vbar,
\eea
where
\bea
\hat g_{11}&=&(1-|v|^2)Dc_0 + (1+|v|^2)De_0 - 4 \Im \!(vDb_0), \non \\
\hat g_{12}&=&Dc_0 - De_0,\non \\
\hat f_1&=&Db_0 + \fft{i}2\vbar (Dc_0-De_0)\,.
\eea
Additionally, note that since we have set the NS three form to zero we are suppressing the
$SL(2,\mathbb R)$ indices from the RR three-form in this truncation.
Using the conventions of \cite{Lu:1998xt}, the metric on the coset
\be
\cM_{\rm hyp}=\frac{SO(4,2)}{SO(4)\times SO(2)} \label{hypermanNS}
\ee
is
\bea
-\frac{1}{8}\Tr dM \w *dM^{-1}&=&\frac{1}{4} \blp d\phi_1^2+d\phi_2^2\brp + \half e^{-\phi_1+\phi_2} dx_1^2 + \half e^{\phi_1}\blp dx_3^2+dx_4^2\brp \non\\
&&+ \half e^{\phi_2} \blp  (d(x_5-x_1x_4)+x_4dx_1)^2+(d(x_6-x_1x_3)+x_3dx_1)^2\brp \non \\
&& + \half e^{\phi_1+\phi_2} \blp dx_2+x_3dx_6+x_4dx_5\brp^2.
\eea
This is related to $\cL_{hyp}$ by the field redefinitions
\bea
\phi-4u_1&=&\phi_2,\non \\
-4u_2 &=& \phi_1,\non \\
\sqrt{2}\, v&=&x_4-ix_,\non3 \\
2\sqrt{2} b_0&=& x_6-x_1x_3 -i(x_5-x_1 x_4),\non \\
c_0- e_0&=& -x_1,\non \\
c_0+ e_0&=& x_2+\half  x_1(x_3^2+x_4^2)\,.
\eea

\subsubsection{Vector multiplet sector}

The scalars in the vector multiplet have kinetic terms
\bea
\cL_{\rm vec}&=& -12 du_3\w *du_3 -\frac{1}{4}d (4u_1+\phi)\w*d (4u_1+\phi).
\label{svecNS}
\eea
We see that the two real scalars in the vector multiplets are $u_3$ and $4u_1+\phi$ and they parameterize the manifold
\be
\cM_v=SO(1,1)\times SO(1,1).
\ee
In terms of $A_1$, $b_1$ and $b_2$, the gauge kinetic terms are
\bea
\cL_{g,kin}&=&-\half e^{-8u_3} F_2\w * F_2 -\half e^{4 u_1-4 u_3+\phi} g_3\w * g_3- \half e^{4u_1+4u_3+\phi} g_2\w *g_2.
\label{LgkinNS}
\eea
We may integrate out the tensor field by dualizing $\hg_3=db_2$ into a vector field.  This is
done  by adding
\be
\Delta \cL=\tb_1\w d\hg_3
\ee
to the Lagrangian. This results in
\bea
\cL_{g,kin}&=& -\half e^{-8u_3} F_2\w * F_2- \half e^{4u_1+4u_3+\phi} g_2\w *g_2-\frac{1}{2} e^{-4 u_1+4 u_3-\phi}  \tg_2 \w * \tg_2,
\eea
along with a Chern-Simons term
\be
\cL_{CS}=  \tg_2 \w b_1 \w F_2,
\ee
where  $\tg_2=d\tb_1$.

The special geometry data for this case is very similar to the Betti-vector sector and is given by the following constrained scalars
\be
X^0 = e^{4u_3}, \qquad X^1 = e^{-2u_1-2u_3-\phi/2}, \qquad X^2 = e^{2u_1-2u_3+\phi/2},
\ee
with $c_{012} = 1$.  The vector field strengths are given by
\be
F^0 = F_2, \qquad F^1 = -g_2, \qquad F^2 = \tg_2.
\ee

\subsubsection{Scalar potential}

The scalar potential has two contributions which we distinguish for clarity:
\bea
\cL_{pot}&=&-\blp V_{gr}+V_{F_{(3)}}\brp, \label{potentialNS}\\
V_{gr}&=&  -12e^{-4u_1-2u_2+2u_3}\blp 1+|v|^2 +e^{4u_2}\brp  +9|v|^2e^{-4u_2+8u_3}  \non \\
&&-2e^{-8u_1-4u_3} \blp e^{4u_2}+e^{-4u_2}(1-|v|^2)^2+2|v|^2\brp, \\
V_{F_{(3)}}&=&\half e^{-4u_1+8u_3+\phi} \Blp 8 |\hat f_0|^2  +e^{4u_2} P^2+e^{-4u_2}  \blp P(|v|^2-1)+4\, \Im\!(\hat f_0 v) \brp^2\Brp,
\eea
where
\be
\hat f_0=f_0-\frac{i}{2} P \vbar.
\ee

\subsection{Fermion Variations}

In order to reduce from $\mathcal N=4$ to $\N=2$, we restrict the transformation parameter
$\varepsilon$. To do this we make the identification
\be
\varepsilon = -i\sigma_2\varepsilon^c,
\ee
where the conjugation is defined by $\varepsilon^c = \gamma_0 C \varepsilon^*$. Additionally we make the same identification for all of the fermions. Given this identification, the components of
\be
\varepsilon = \btop{\varepsilon_1}{\varepsilon_2}
\ee
satisfy the symplectic-Majorana condition, $\varepsilon_1 = - \varepsilon_2^c,$ which can be expressed as
\be
\varepsilon^i \equiv \epsilon^{ij}\varepsilon_j = \varepsilon_i^c.
\ee
We can then identify the two hyperini as
\bea
\zeta^1 &=& \lambda - 2(\psi^{(5)} + \psi^{(7)}),\non \\
\zeta^2 &=& \psi^{(5)} - \psi^{(7)},
\eea
and the gaugini are given by some linear combination of
\bea
\chi^1 &=& \lambda + 2(\psi^{(5)} + \psi^{(7)}),\non \\
\chi^2 &=& \psi^{(5)} + \psi^{(7)} +2\psi^{(9)}.
\eea

The $\N=2$ susy transformations are then
\bea
\delta\zeta^1 &=& \Big[-\ft{i}2\gamma\cdot\partial(\phi-4u_1) + \ft12e^{\phi/2-2u_1}(e^{-2u_2}(1-|v|^2)+e^{2u_2})(\gamma\cdot g_1 + 2ie^{-2u_1-2u_3-\phi/2}) \non \\
&&\phantom{\Big[} +\ft12e^{\phi/2-2u_1}(e^{-2u_2}(1+|v|^2)-e^{2u_2})(\gamma\cdot h_1 - iPe^{4u_3}) \non \\
&& \phantom{\Big[}+ i e^{\phi/2-2u_1-2u_2}[v(\gamma\cdot f_1 - ie^{4u_3}f_0) - \bar{v}(\gamma\cdot \bar f_1 - ie^{4u_3}\bar f_0)]\Big]\varepsilon  \non \\
&&-\bar v e^{\phi/2-2u_1}\Big[(\gamma\cdot g_1 + 2ie^{-2u_1-2u_3-\phi/2}) - (\gamma\cdot h_1 - iPe^{4u_3}) + 2i(\gamma\cdot f_1 - ie^{4u_3}f_0)\Big]\varepsilon^c, \non \\
\delta\zeta^2 &=& \Big[-i\gamma\cdot\partial u_2 - \ft14e^{\phi/2-2u_1}(e^{-2u_2}(1-|v|^2)+e^{2u_2})(\gamma\cdot g_1 + 2ie^{-2u_1-2u_3-\phi/2}) \non \\
&& \phantom{\Big[}-\ft14e^{\phi/2-2u_1}(e^{-2u_2}(1+|v|^2)-e^{2u_2})(\gamma\cdot h_1 - iPe^{4u_3}) \non \\
&&\phantom{\Big[} - \ft{i}2 e^{\phi/2-2u_1-2u_2}[v(\gamma\cdot f_1 - ie^{4u_3}f_0) - \bar{v}(\gamma\cdot \bar f_1 - ie^{4u_3}\bar f_0)]\Big]\varepsilon \nonumber\\
&&\phantom{\Big[}+ \ft{i}2e^{2u_2}(\gamma\cdot \overline{Dv} + 3\bar v e^{4u_3})\varepsilon^c,\non \\
\delta\chi^1 &=& \Big[ -\ft{i}2\gamma\cdot\partial(\phi+4u_1) -\ft{1}{12} e^{-\phi/2-2u_1+2u_3}\gamma\cdot \tilde g_2 - \ft14 e^{\phi/2+2u_1+2u_3}\gamma\cdot g_2 \non  \\
&& \phantom{\Big[}- ie^{-4u_1-2u_3}(e^{-2u_2}(1-|v|^2) + e^{2u_2})\Big]\varepsilon + 2i\bar ve^{-4u_1-2u_3}\varepsilon^c, \non \\
\delta\chi^2 &=& \Big[3i\gamma\cdot\partial u_3 + \ft14e^{-4u_3}\gamma\cdot F_2  -\ft{1}{24}e^{-\phi/2-2u_1+2u_3}\gamma\cdot \tilde g_2 + \ft{1}{8}e^{\phi/2+2u_1+2u_3}\gamma\cdot g_2 \nonumber \\
&& + \ft{i}2\Big(e^{-4u_1-2u_3}(e^{-2u_2}(1-|v|^2) + e^{2u_2}) - 6e^{4u_3} \non \\
&& \phantom{\ft{i}2\Big(} + e^{\phi/2-2u_1+4u_3}[e^{-2u_2}((1+|v|^2)P + 2ivf_0 - 2i\bar v \bar f_0) - e^{2u_2}P]\Big)\Big]\varepsilon \nonumber \\
&& - \Big[\bar ve^{-4u_1-2u_3} - 3\bar ve^{-2u_2+4u_3} + 2(f_0 - \ft{i}2\bar v P)e^{\phi/2-2u_1+4u_3}\Big]\varepsilon^c,
\eea
along with
\bea\label{eq:NSgrav}
\delta\psi_\alpha
&=&\Big[D_\alpha +\ft{i}{24}(\gamma_\alpha{}^{\beta\gamma}
-4\delta_\alpha^\beta\gamma^\gamma)\Big(e^{-4u_3}F_{\beta\gamma} + e^{-\phi/2-2u_1+2u_3}\tilde g_{2\,\beta\gamma} - e^{\phi/2+2u_1+2u_3}g_{2\,\beta\gamma}\Big) \nonumber \\
&&+\ft16\gamma_\alpha\Big(e^{-4u_1-2u_3}(e^{-2u_2}(1-|v|^2)+e^{2u_2})+3e^{4u_3} \nonumber\\
&&\phantom{+\gamma_\alpha\Big(} - \ft1{2}e^{\phi/2-2u_1+4u_3}(e^{-2u_2}((1+|v|^2)P + 2ivf_0 - 2i\bar v \bar f_0) - e^{2u_2}P)\Big)\Big]\varepsilon \nonumber \\
&& + \ft{i}6\gamma_\alpha\Big[(2e^{-4u_1-2u_3}+3e^{-2u_2+4u_3})i\bar v - 2e^{\phi/2-2u_1+4u_3}(f_0 - \ft{i}2\bar vP)\Big]\varepsilon^c.
\eea
%

\section{Field redefinitions and conventions}\label{app:map}

Here we make explicit the relations between our reduction ansatz and those presented in Refs.~\cite{Bena:2010pr} and \cite{Liu:2011dw}. Our ansatz follows the conventions of Ref.~\cite{Bena:2010pr} for the metric and the five-form ansatz. However, we have chosen a manifestly $SL(2,\mathbb R)$ covariant notation for the three-form.  Our three-form ansatz is related to that of Ref.~\cite{Bena:2010pr} according to
\bea
&& b_2^1 = B_2 + \ft12 b F_2, \qquad b_1^1 = B_1, \qquad 3ib_0^1 = M_0,\kern2.9em
 \qquad c_0^1 = b, \qquad e_0^1 = \tilde b, \qquad j_1^1 = 0,\non  \\
&&  b_2^2 = C_2 + \ft12 c F_2, \qquad b_1^2 = C_1, \qquad  3ib_0^2 =
 N_0 + a M_0, \qquad  c_0^2 = c, \qquad  e_0^2 = \tilde c,\qquad
 j_0^2 = P. \non\\
\eea

Additionally, the conventions here are consistent with that of the three-form in Ref.~\cite{Liu:2011dw}. But for the metric and five-form the relations are given by
\bea
&&u_1 = \fft12(B_1+B_2), \qquad u_2 = \fft12(B_1-B_2), \qquad u_3 =
-\fft16(B_1+B_2)-\fft13C, \qquad v = \alpha, \non \\
&& e^Z = 4+\phi_0, \qquad K_1 = \mathbb A_1, \qquad K_{21} = p_{21},
\qquad K_{22} = p_{22}.
\eea

\providecommand{\href}[2]{#2}\begingroup\raggedright\endgroup

\end{document}